\documentclass[10pt,aps,prx,twocolumn,superscriptaddress,longbibliography]{revtex4-2} 
\pdfoutput=1
\usepackage{graphicx}  
\usepackage{dcolumn}   
\usepackage{bm}        
\usepackage{amssymb}   
\usepackage{color}
\usepackage{float}
\usepackage{subcaption}
\usepackage[utf8]{inputenc}
\usepackage{braket}
\usepackage{amsmath}
\usepackage{upgreek}

\usepackage[linkcolor=blue,colorlinks=true,urlcolor=blue,citecolor=red]{hyperref}
\usepackage[nameinlink]{cleveref}

\renewcommand{\figurename}{Fig.}
\usepackage{textgreek}
\newcommand{\vac}{V_{ac}}
\newcommand{\vdc}{V_{dc}}
\newcommand{\vtot}{V_{tot}}
\newcommand{ \micro}{\textmu}
\newcommand{ \ep}{\varepsilon_\pm}
\DeclareMathOperator{\atantwo}{atan2}
\usepackage{ulem}

\begin{document}

\crefname{section}{Sec.}{Secs.}

\title{Bichromatic synchronized chaos in coupled optomechanical nanoresonators}

\author{Guilhem Madiot}
\affiliation{Centre de Nanosciences et de Nanotechnologies, CNRS,  Université Paris-Saclay, Palaiseau, France}

\author{Franck Correia}
\affiliation{Centre de Nanosciences et de Nanotechnologies, CNRS,  Université Paris-Saclay, Palaiseau, France}

\author{Sylvain Barbay}
\affiliation{Centre de Nanosciences et de Nanotechnologies, CNRS,  Université Paris-Saclay, Palaiseau, France}

\author{Rémy Braive}
\affiliation{Centre de Nanosciences et de Nanotechnologies, CNRS,  Université Paris-Saclay, Palaiseau, France}
\affiliation{Université de Paris, F-75006 Paris, France}

\begin{abstract}

Synchronization and chaos are two well known and ubiquitous phenomena in nature. Interestingly, under specific conditions, coupled chaotic systems can display synchronization in some of their observables. Here, we experimentally investigate bichromatic synchronization on the route to chaos of two non-identical mechanically coupled optomechanical nanocavities. 
Electromechanical near-resonant excitation of one of the resonators evidences hysteretic behaviors of the coupled mechanical modes which can, under amplitude modulation, reach the chaotic regime. The observations, allowing to measure directly the full phase space of the system, are accurately modeled by coupled periodically forced Duffing resonators thanks to a complete calibration of the experimental parameters.
This shows that, besides chaos transfer from the mechanical to the optical frequency domain, spatial chaos transfer between the two nonidentical subsystems occurs. Upon simultaneous excitations of the coupled membranes modes, we also demonstrate bichromatic chaos synchronization between quadratures at the two distinct carrier frequencies of the normal modes. Their respective quadrature amplitudes are consistently synchronized thanks to the modal orthogonality breaking induced by the nonlinearity. 
 Meanwhile, their phases show complex dynamics with imperfect synchronization in the chaotic regime. Our generic model agrees again quantitatively with the observed synchronization dynamics. These results set the ground for the experimental study of yet unexplored collective dynamics of e.g synchronization in arrays of strongly coupled, nanoscale  nonlinear oscillators for applications ranging from precise measurements to multispectral chaotic encryption and random bit generation, and to analog computing, to mention a few.

\end{abstract}

\maketitle

\section{INTRODUCTION}

Synchronization is a common phenomenon where an oscillating physical quantity tends to develop a phase preference when coupled to a drive or to another oscillating system \cite{PIK01}. Since Huygens well-known preliminary studies of synchronization between two mechanically coupled pendulum clocks \cite{Huygens}, such phenomenon is ubiquitous in many fields of sciences e.g. in physics \cite{Rosenblum1997,Pisarchik2005,ZhangPRL2012,ZhangPRL2015}, chemistry \cite{Yam1976}, biology \cite{Glass2001}, ecology \cite{Blasius1999}, economy \cite{Volos2012} and even in sociology \cite{Moussaid2009}. Although it may look like a unique phenomenon, synchronization features may occur within different dynamical regime in which the system is set. Among these regimes, chaotic dynamics, interdisciplinary by essence, has attracted a lot of attention. Even though synchronization of chaotic
system may seem counterintuitive \cite{Pecora1990}, the investigation and implementation of such a phenomenon is particularly meaningful and has applications from large scale systems as in meteorologic
\cite{DuanePRL,Duane} and atmospheric general circulation \cite{Hiemstra,Lunkeit} models, to small scales with atomic clocks \cite{Patra}. Beyond this, chaos in arrays of synchronized systems is of great interest for generating random numbers \cite{shore2015} potentially used for robust and encrypted communications in optics \cite{CuomoIEEE,Argyris2005,Annovazzi-Lodi,Mirasso} and astronomy \cite{MARINHO2005230}. It could also be relevant in artificial neuron networks model \cite{Hsu,Milanovic}, ultra-precision measurements \cite{Fiderer2018}, fundamental tests on the physical conditions for classical dynamics \cite{PhysRevLett.114.013601}.

Signatures of synchronization can be imprinted on the dynamics of both quadrature of the investigated signal, i.e. the amplitude and the phase. However, this latter exhibits a much richer dynamic which may be evidenced in time by periodic, erratically distributed jumps and many others (as described in Kuramoto model \cite{Yam1976}). Imperfect phase synchronization \cite{ParkPRE} refers to the case where intermittent phase slips occur while the amplitude is still synchronized. This phenomenon is barely studied experimentally although it constitutes an universal paradigm anchored in a wide range of disciplines such as physics \cite{Cross2003,BlackburnPRB}, electronics \cite{Wu1995}, and even neurosciences \cite{SHUAI1999289} to name a few. A complete understanding of these concepts can be reached with a classical model of coupled driven resonators. Therefore, the dynamics resulting from this basic model can be applied and is also relevant to many other fundamental studies such as spatio-temporal phase transition physics \cite{Clerc2018,Martens10563,pelka2019chimera}, patterns formations \cite{Marquardt2015} or synchronized neutrinos oscillations \cite{Mirizzi2016} encountered in particle physics. In this frame, mechanical and optomechanical systems constitute a platform of choice thanks to their experimental adaptability to host most of fundamental concepts of nonlinear dynamics. Many nonlinear dynamical effects have been demonstrated in single or coupled Nano ElectroMechanical Systems (NEMS) \cite{Midolo2018,Gao_2019,Unterreithmeier2009,Eichler2012,Gajo2017,Okamoto2013,chowdhuryPRL}, including synchronization \cite{Shim95,DongAPL,MathenyPRL} and chaos \cite{KarabalinPRB}. In line with these studies, chaos with optomechanical systems has been extensively studied theoretically \cite{PhysRevA.90.043839,PhysRevA.84.021804,Jin2017}. Only recently and using a single resonator, experimental demonstrations of chaos in optomechanical cavity have been reported \cite{Navarro-Urrios2017,Wu2017} with a strong optical driving.	Here, thanks to a forcing amplitude modulation technique, we emphasize mechanical chaos using two mechanically coupled optomechanical cavities. Beyond chaos transfer from mechanics to optics, we evidence energy transfer between two non-identical mechanical resonators through the spring coupling. Despite their intrinsic natural frequencies mismatch, chaos is simultaneously generated at two distinct carrier frequencies opening potential new avenues to synchronized multifrequency data encryption \cite{CuomoIEEE,Argyris2005,Annovazzi-Lodi,Mirasso} and random number generation \cite{shore2015}.
Interestingly, the quadrature amplitudes of the two chaotic signals at two different tones are synchronized.
In parallel the quadrature phases evidence different regimes from phase synchronization to imperfect phase synchronization through phase desynchronization.
The unequivocal description of these regimes is enabled by the measurement method which give a direct access to the dynamical variables, without making any use of reconstructed signals.
These phenomena can be understood and fully supported by numerical simulations based on a classical model using Duffing resonators, which beyond nanomechanics, can be used in many fields including superconducting Josephson amplifier \cite{ROY2016740}, ionization plasma \cite{HsuanAPL} and complex spatiotemporal behaviors such as chimera sates \cite{Clerc2018}.

Our study relies on the mechanical coupling between two non-identical optomechanical systems: electro-capacitively driven Fabry-Pérot cavities which are described in \cref{sec2}. After a preliminary study of their linear mechanical properties in \cref{sec3:a} allowing to extract useful modeling parameters, a strong resonant driving field is applied and evidences hysteretic behavior in \cref{sec3:b}. As such, we experimentally demonstrate in \cref{sec4} how low-frequency amplitude modulation exerted on driven coupled Duffing resonators can boost the nonlinearity in the material so strongly that it leads to a period-doubling cascade and then to chaos. Experimental bifurcation diagrams are reconstructed from the measured time traces when either eigenmode is driven. In each configuration, the largest Lyapunov exponent is computed attesting the chaotic behaviors. When simultaneously driven, the eigenmode orthogonality breaking due to the Duffing nonlinearity is shown in \cref{sec5} to allow them to couple. We then investigate the synchronization of the eigenmode amplitude and phase responses. While the amplitude responses are found to be robustly synchronized, we evidence several phase synchronization regimes including complete synchronization, desynchronization and imperfect synchronization. The experimental results are in good quantitative agreement with the proposed model. In \cref{sec6} we conclude.

\section{EXPERIMENTAL SYSTEM AND MEASUREMENT SCHEME} \label{sec2}

\begin{figure}[htp]
\includegraphics[scale=0.5]{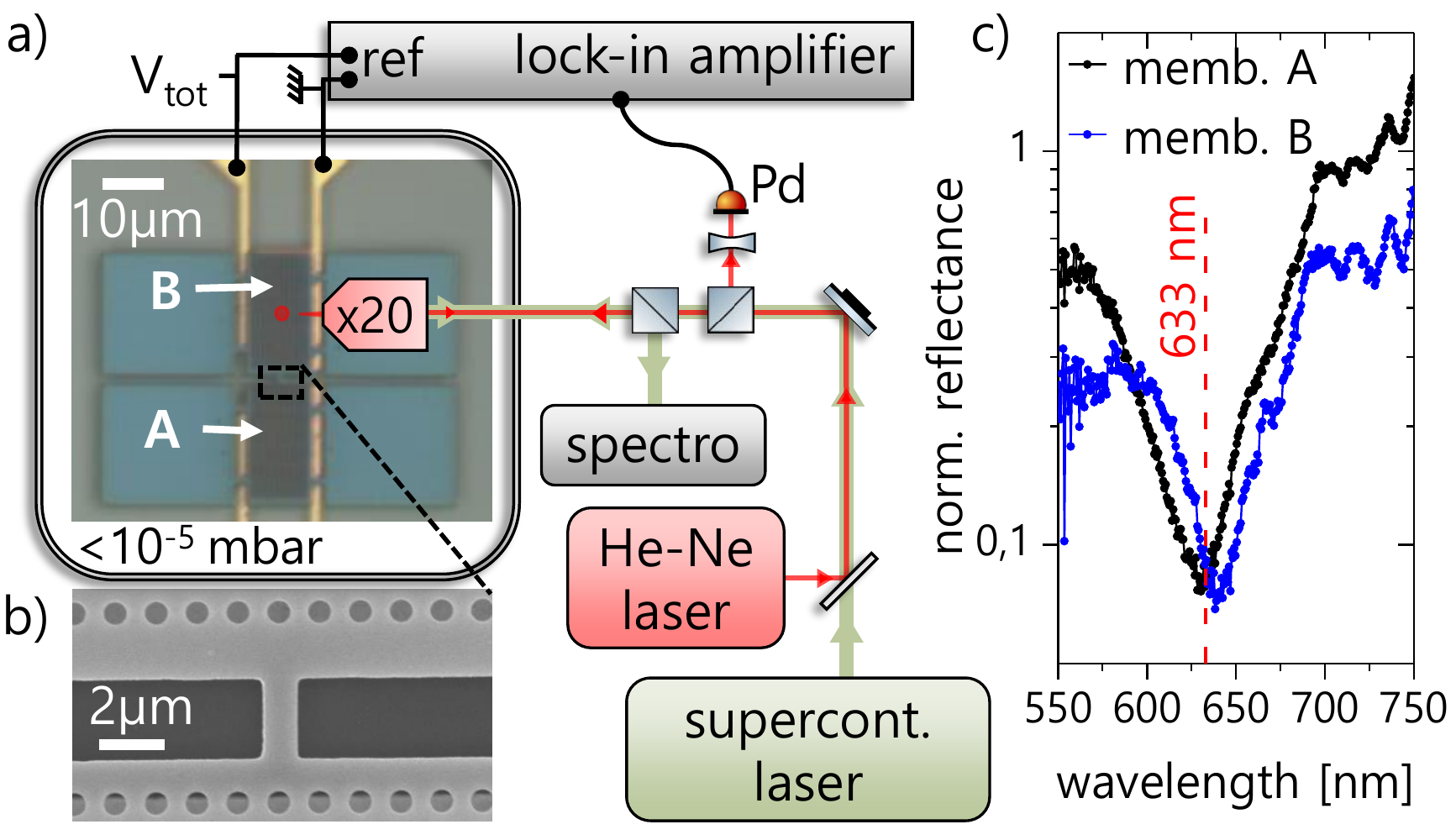}
{\phantomsubcaption\label{fig1:a}}
{\phantomsubcaption\label{fig1:b}}
{\phantomsubcaption\label{fig1:c}}
\caption{a) excitation and detection setup including an optical microscope image of the integrated system. Two suspended membranes are attached by four mesas and bridged by a coupling nanobeam. Gold stripes are visible underneath the membranes for electro-actuation. b) SEM micrograph of the coupling beam. c) Reflectance spectra for the two Fabry-Pérot cavities formed by membranes A (black) and B (blue) and the underneath substrate.}
\end{figure}

The experimental system (see \cref{fig1:a}) consists of two coupled mechanical micro-resonators made of a common 260 nm thick InP layer suspended over a 380 nm air gap. Each membrane is a 10$\times$20 \micro m$^2$ rectangle pierced with a square lattice of cylindrical holes. This array both permits to reduce the mechanical masses, increasing the mechanical resonator natural frequencies and allows an enhancement of the out-of-plane reflectivity \cite{Antoni:11}. The two membranes are mechanically coupled through a 1 \micro m wide and 1.5 \micro m long bridge (see \cref{fig1:a,fig1:b}). A pair of gold interdigitated electrodes (IDEs) is positioned on the substrate below each resonator and allow for independent actuation of the mechanical resonators. The fabrication process is described in \cite{chowdhuryAPL}. All measurements are done at room temperature and the chip is placed in a vacuum chamber pumped below $10^{-5}$ mbar. 

The system composed of the membrane plus the IDEs constitutes a low-finesse optical cavity which we can probe for measuring the mechanical displacement of the suspended membranes. 
The reflectance spectra shown in \cref{fig1:c} are measured by means of a supercontinuum laser reflected on the centers of the at-rest membranes. The beam is focused down to 5 \micro m on the membrane center with a $\times$20-microscope objective to inject the cavity. The resulting spectra are normalized with a reference measurement obtained by pointing the laser at a gold planar surface available on the chip. The normalized reflectance shows a pronounced dip typical of a Fabry-Pérot cavity resonance and centered around 630 nm with an optical Q-factor of about 10. The reflectance dip matches with the Helium-Neon wavelength $\lambda=633$ nm which is used for the optical readout the membrane displacements induced by the capacitive field generated by the IDEs when submitted to a voltage $\vtot$.
 The electro-capacitive force exerted on one membrane reads \cite{Unterreithmeier2009}: 
\begin{equation}
F = -\frac{1}{2}\frac{dC}{dx}\vtot^2 
\label{EMforce}
\end{equation}
 where $C(x)$ is the position-dependent capacitance of the membrane/IDEs system. The driving voltage is $\vtot=\vdc+\vac \cos(\omega_d t)$ where $\vdc$ is a static voltage and $\vac$ is the amplitude of the AC driving at frequency $\omega_d$. Each membrane can be independently probed. A photodetector converts the reflected optical field into an electrical signal sent to a lock-in amplifier to access the demodulation amplitude $\eta R_{A,B}$ and phase $\theta_{A,B}$ where A or B here refer to the probed resonator and $\eta$ converts a given measured voltage to the corresponding mechanical displacement $R_{A,B}$. The calibration of $\eta$ can be obtained thanks to a Michelson interferometer and we estimate the conversion constant $\eta\approx0.5$ mV.nm$^{-1}$ (see \cref{appen_calib}). The RF input is demodulated with a passband filter centered at $\omega_d$ enabling a higher signal-to-noise ratio and the demodulator internal frequency is locked to the applied harmonic excitation of the IDEs.

\section{SINGLY DRIVEN COUPLED RESONATORS}
\subsection{Linear Regime}\label{sec3:a}

\begin{figure*}
\centering
\includegraphics[scale=0.54]{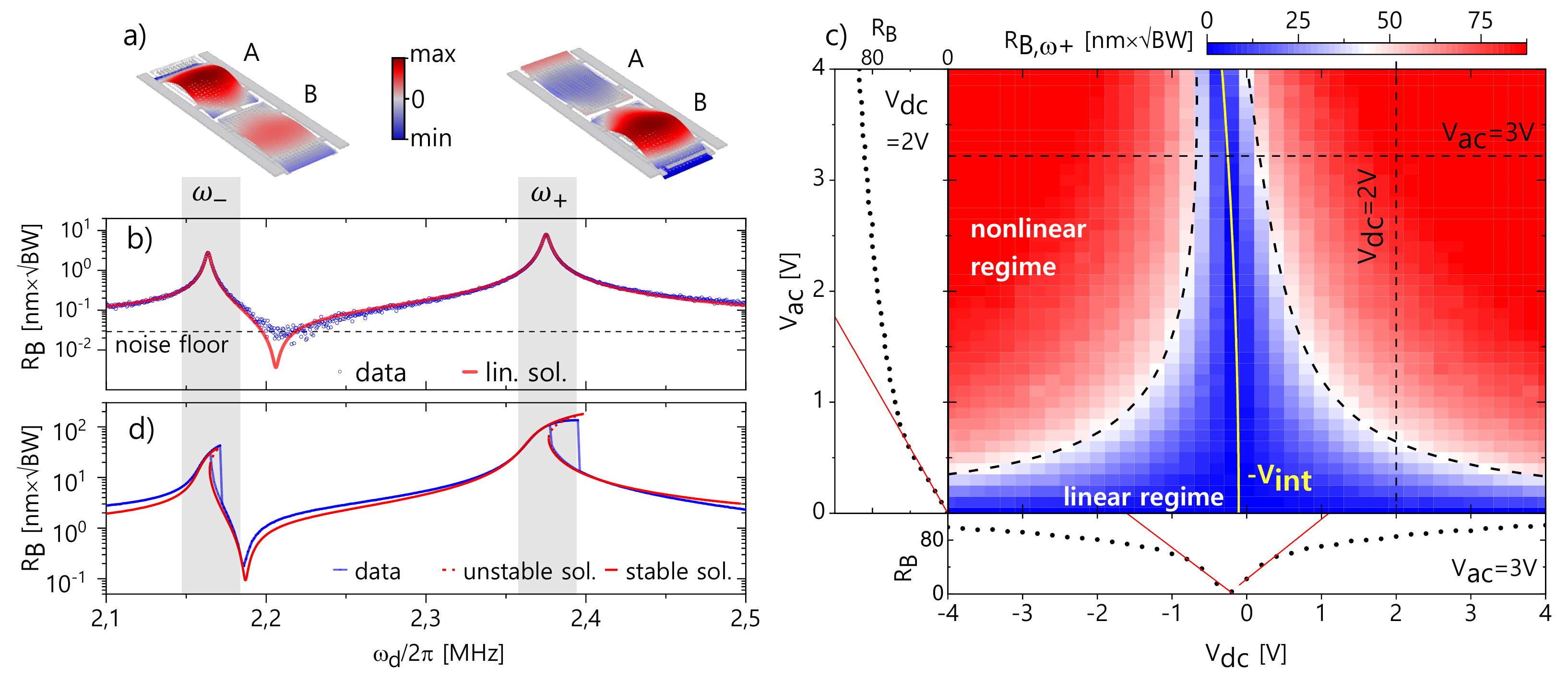}
{\phantomsubcaption\label{fig2:a}}
{\phantomsubcaption\label{fig2:b}}
{\phantomsubcaption\label{fig2:c}}
{\phantomsubcaption\label{fig2:d}}
\caption{a) FEM simulation of the out-of-plane displacement field for the symmetrical ($\omega_-$) and the antisymmetrical ($\omega_+$) eigenmode taking into account the frequency mismatch. b) Spectral response of membrane B when driven with $\vac=1$V and $\vdc=0.5$V. Experimental data (blue points) are fitted with a linear model (red line). The detection noise level floor is marked by a black dashed line. c) Membrane B experimental resonant response amplitude in ($\vac$,$\vdc$) parameter space for $\omega_d=\omega_+$. Horizontal and vertical slices are represented for respectively $\vac=3$V and $\vdc=2$V with a linear fit (red line) in the non saturated regime. The iso-$\vdc^\mathrm{eff}\vac$ curves (black dashed) separate the linear and Duffing-Duffing regimes. The internal stress (yellow line) parabolically shifts with $\vac^2$. d) Spectral response of membrane B at $\vac=3$V and $\vdc=2$V. Data are recorded with upward and downward $\omega_d$ sweeps (blue symbols) and are fitted with the Duffing-Duffing model with stable (red line) and unstable (red dotted) solutions. All displacement measurement are performed with a domulation $\mathrm{BW}=100$Hz.}
\end{figure*}

To build the model describing the system, we first introduce the potential energy associated to uncoupled harmonic resonators:
\begin{equation}
U_\mathrm{lin}(x_A,x_B) = \frac{1}{2}(\omega_{A,0}^2x_A^2 + \omega_{B,0}^2x_B^2)
\end{equation}

with $\omega_{A,0}$ (resp. $\omega_{B,0}$) the natural frequency of the mode resonator A (resp. B). The mechanical beam joining the two membranes contributes to the dynamics through the coupling spring constant G. This results in a coupling potential $U_\mathrm{coup} (x_A,x_B )=Gx_A x_B + \frac{G}{2}(x_A^2+x_B^2)$ with the resonators self-coupled frequencies $\omega_A$ and $\omega_B$ such that $\omega_A^2=\omega_{A,0}^2+G$ and $\omega_B^2=\omega_{B,0}^2+G$.
 We consider mass normalized physical quantities in our model. We introduce linear mechanical damping terms $\Gamma_A$ and $\Gamma_B$ for each resonator and describe the problem with the following coupled master equations \cite{Zanette_2018}:
\begin{equation}
\left\{
\begin{aligned}
\ddot{x}_A + \Gamma_A \dot{x}_A + \frac{\partial U_\mathrm{tot}}{\partial x_A} =& 0  \label{MasterEq}\\
\ddot{x}_B + \Gamma_B \dot{x}_B + \frac{\partial U_\mathrm{tot}}{\partial x_B} =& F_B\cos(\omega_dt) 
\end{aligned}
\right.
\end{equation}

where the total potential energy is $U_\mathrm{tot}=U_\mathrm{lin}+U_\mathrm{coup}$. Note that the forcing term $F_B$ is the only non-zero right-hand side term because we only excite the membrane B. 
 The stationary solutions of \cref{MasterEq} can be derived by assuming an oscillatory solution $x_{A,B}=r_{A,B}  \cos(\omega_d t+\theta_{A,B})$ where $r_{A,B}$ is the mechanical displacement of each membrane. 
 
In this first experiment we will characterize the mechanical response of the system in the linear regime. Membrane B is driven $\vdc = 0.5$V and $\vac = 1$V to the corresponding set of IDEs. The He-Ne laser is focused on membrane B so that we record its response amplitude $R_B$ while the driving frequency $\omega_d$ is swept. The driven mechanical system exhibits a large variety of normal modes ranging from 1 to 10 MHz. We focus our attention on the lowest frequency modes corresponding to the coupled fundamental modes of the membranes. The modes centered respectively at $\omega_- = 2\pi\times2.161$ MHz and $\omega_+=2\pi\times2.369$ MHz are identified as the symmetrical ($-$) and antisymmetrical ($+$) normal modes. 

By fitting the theoretical response of resonator B (\cref{fig2:b}) we obtain the self-coupled frequencies $\omega_A=2\pi\times2.187$ MHz, $\omega_B=2\pi\times2.345$ MHz, and the dampings $\Gamma_A=2\pi\times2.4$ kHz and $\Gamma_B= 2\pi\times4.3$ kHz. The frequency mismatch arises because of fabrication imperfections. The mechanical quality factors of the normal modes can be computed and are of the order of 660. The normal mode splitting expected for coupled identical resonators is found to be $G/\omega_B\approx2\pi\times130$ kHz. The mode coupling is further attested on by the presence of a destructive interference dip around 2.21 MHz which is typical of a Fano resonance \cite{Joe_2006,Limonov2017,Stassi2017} between the nearly identical resonators. In the linear regime and for a low driving amplitude, the Fano dip minimum is below the detection noise floor but its presence on the spectrum is nevertheless clearly visible.

The constitutive resonator frequency difference leads the normal mode at $\omega_-$ (resp. $\omega_+$) to be dominated by the motion of membrane A (resp. B). This conclusion is confirmed by measuring $\omega_B-\omega_A$ with a dielectric tuning technique described in \cref{appen:a}. Altogether it implies that $\omega_-\approx\omega_A$ and $\omega_+\approx\omega_B$. Finite Elements Method (FEM) simulations confirm the normal modes respective displacement fields and the energy imbalance due to a natural frequency mismatch (\cref{fig2:a}). 
 The amplitude of the normal mode ($+$) is almost 4 times higher than amplitude of mode ($-$) due to this imbalance. 
  In the context of identical resonators, the strong coupling regime is established when the criterion $G/\omega_A >\Gamma_A$ is satisfied \cite{Zanotto2018}.
This criterion applies in our experiment, however, since the resonator frequency mismatch is about twice as large as the minimum normal mode splitting we are rather in an intermediate regime between the strong and weak coupling cases.

When expanded, the expression for the electrocapacitive force \cref{EMforce} includes a static component $\propto(\vdc^2+ \vac^2/2)$ that displaces the resonator by a negligible offset plus an off-resonant term at frequency $2\omega_d$ which is ignored in our model. A measurement of the displacement amplitude at demodulation frequency $2\omega_d$ indeed reveals an amplitude response less than 3\% of the one of the driven mode amplitude at $\omega_{d}$. Therefore, the driving force amplitude can be related to the experimental parameters with $F_B=|\frac{1}{m_\mathrm{eff}}  \frac{dC}{dx} \vdc \vac |$ where $m_\mathrm{eff}$=186 pg is the effective mass computed at fundamental eigenfrequency of normal mode (+) by FEM.
 The resonant amplitude $R_B(\omega_d=\omega_+)$ is mapped over the parameter space \{$\vdc$,$\vac$\} as shown in \cref{fig2:c}.
 The stationary solutions of \cref{MasterEq} indicate that the resonators responses are both linear with the strength $F_B$. 
 This linear dependence is independently checked (red lines in \cref{fig2:c}) with varying $\vdc$ (by line) or $\vac$ (by column) both allowing the electro-capacitive force to be consistently calibrated: $dC/dx\approx2.2$ \micro N/V$^2$.

The model takes into account a small offset in the effective static voltage $\vdc^\mathrm{eff}=\vdc+V_\mathrm{int}$ where $V_\mathrm{int}$ corresponds to the internal stress of the membrane. We observe a small shift of the internal stress with increasing $\vac$ due to the dielectric tuning induced by the static component $\vac^2$ \cite{RiegerAPL}.

\subsection{Nonlinear Regime}\label{sec3:b}
In \cref{fig2:c}, a domain of saturation of the mechanical response settles when the product $\vdc^\mathrm{eff} \vac$ is greater than $1.3$V$^2$. The corresponding iso-$\vdc^\mathrm{eff}\vac$ curves delimit a threshold between the linear and the nonlinear regimes. They are fitted (dashed black lines) using the data points at the frontier between the two domains.

The system response in the nonlinear regime is shown on \cref{fig2:d} for $\vdc=2$V and $\vac$=3V and displays two histeretic regions around $\omega_-$ and $\omega_+$ that are evidenced by sweeping $\omega_d$ forward and backward. This saturation arises from intrinsic mechanical nonlinearities \cite{RhoadsJDS} which can be modeled thanks to a Duffing oscillator model \cite{chowdhuryAPL,chowdhury2019weak}. 

We follow the same approach as in \cref{sec3:a} and introduce anharmonicity to the uncoupled harmonic resonators potential energy through the nonlinearity $\beta$:
\begin{equation}\label{nlMasterEq}
U_\mathrm{NL} = U_\mathrm{lin} + \frac{1}{4}\beta[x_A^4 + x_B^4]
\end{equation} 
 The nonlinear dynamics is still governed by \cref{MasterEq} but considering the new total potential energy $U_\mathrm{tot}=U_\mathrm{coup}+U_\mathrm{NL}$. The stationary solutions for the resonators amplitudes $r_A$ and $r_B$ and phases $\theta_A$ and $\theta_B$ can be described by the following set of equations (see details in \cref{appen:b}):

\begin{equation}
\label{NLsys}
\left\{
\begin{aligned}
\dot{r}_A =& \frac{-\gamma_A}{2} r_A + \frac{g}{2} r_B \sin(\theta_A-\theta_B)\\
\dot{r}_B =& \frac{-\gamma_B}{2} r_B - \frac{g}{2} r_A \sin(\theta_A-\theta_B) \\ &+ \frac{f_B}{2}\sin(\theta_B)\\
r_A\dot{\theta}_A =& \frac{-r_A}{2} \left[ 2(\delta-\Delta\omega) + \frac{3}{4}\tilde{\beta}  r_A^2\right] \\ &+ \frac{g}{2} r_B \cos(\theta_A-\theta_B)\\
r_B\dot{\theta}_B =& \frac{-r_B}{2} \left[ 2\delta + \frac{3}{4}\tilde{\beta}  r_B^2 \right] \\ &+ \frac{g}{2} r_A \cos(\theta_A-\theta_B) + \frac{f_B }{2}\cos(\theta_B)\\
\end{aligned}
\right.
\end{equation}

  where we define the normalized forcing strength $f_B=F_B/\omega_d^2$, dampings $\gamma_i=\Gamma_i/\omega_d$, coupling $g=G/\omega_d^2$, frequency mismatch $\Delta\omega=(\omega_B-\omega_A)/\omega_d$, detuning $\delta=(\omega_B-\omega_d)/\omega_d$ , Duffing nonlinearity $\tilde{\beta}=\beta/\omega_d^2$ and rescaled time $\omega_d^{-1}$. The variables $r_A$ and $r_B$ are left in units of nanometer to compare with the experimental results.

  By assuming the permanent regime $\dot{r}_A=\dot{r}_B=\dot{\theta}_A=\dot{\theta}_B=0$, in \cref{nlMasterEq}, the frequency domain response amplitudes $r_B$ is numerically solved using the experimental parameters extracted from \cref{fig2:b} and previously discussed. We adjust the solution on the experimental data by fitting with the remaining free parameter $\beta$ and extract $\beta=(2\pi)^2\times6.71\times10^{-6}$ MHz$^2$.nm$^{-2}$. The resulting curve shown in \cref{fig2:d} is composed of a stable solution (red line) and an unstable solution (red dashed line). It has been checked that another model including only one anharmonic resonator coupled with a harmonic resonator does not permit to describe the double bistability we experimentally observe. 
We conclude from the good agreement of our model with the experiment that a linear spring coupling satisfactorily describes the mechanical interaction between the membranes.

\section{CHAOTIC DYNAMICS UNDER AMPLITUDE MODULATION}\label{sec4}

\begin{figure*}
\centering
\includegraphics[scale=0.37]{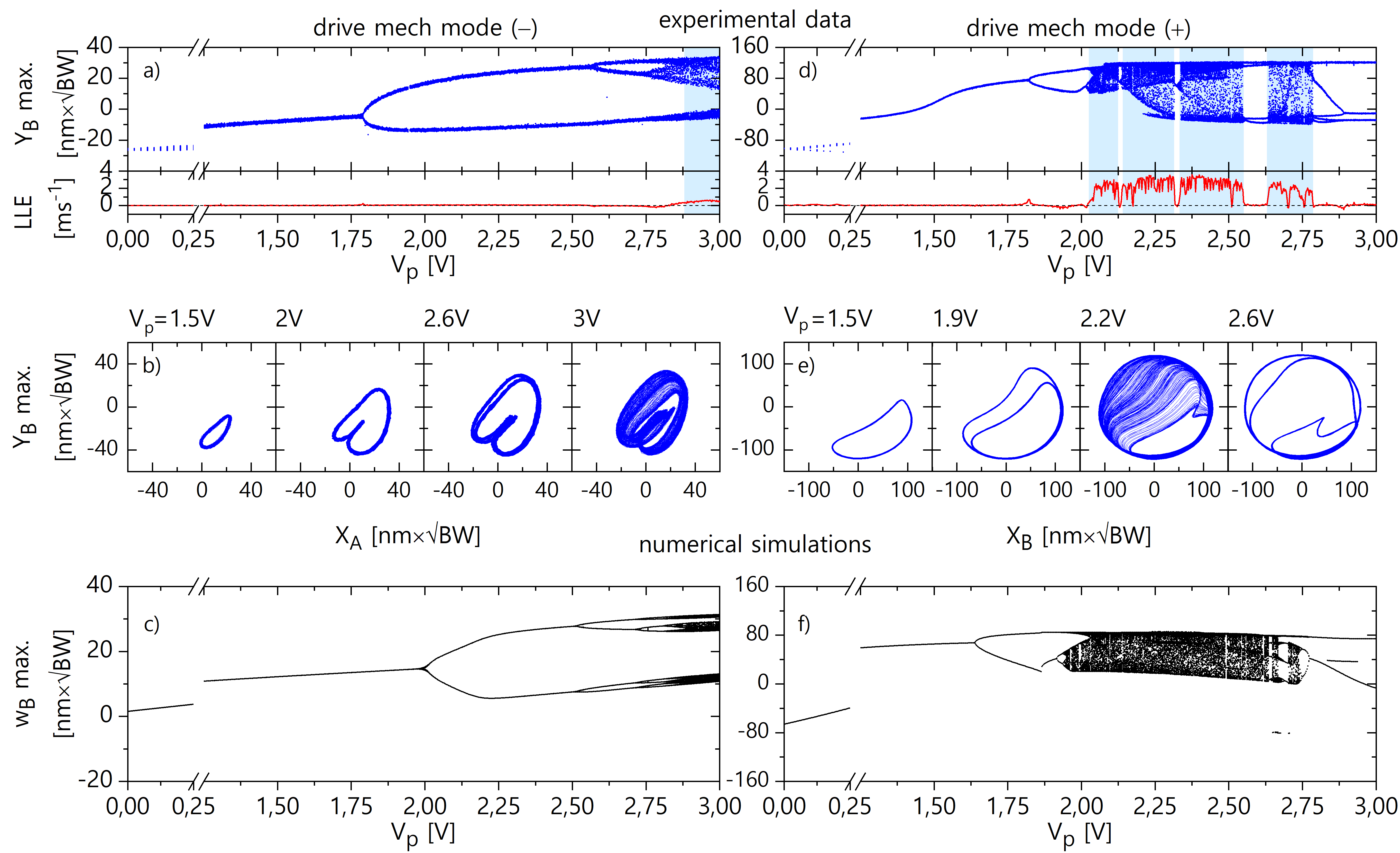}
{\phantomsubcaption\label{fig3:a}}
{\phantomsubcaption\label{fig3:b}}
{\phantomsubcaption\label{fig3:c}}
{\phantomsubcaption\label{fig3:d}}
{\phantomsubcaption\label{fig3:e}}
{\phantomsubcaption\label{fig3:f}}

\caption{Experimental and numerical bifurcation diagrams under single driving and by reading the displacement of membrane B. Measurement and simulations are performed by driving either the symmetrical (left column $\omega_d=2\pi\times2.164$ MHz) or anti-symmetrical resonance (right column $\omega_d=2\pi\times2.379$ MHz) with $\vdc=2$V, $\vac$=3Vand $\omega_p=2\pi\times7$ kHz. a)-d) experimental bifurcation diagrams built by sweeping $V_p$ and reading membrane B with the associated largest Lyapunov exponent (LLE). Note the broken axis. b)/e) Phase portraits at different dynamical regimes. c)/f) Numerical simulation of bifurcation diagrams built from the maxima of quadrature $w_B$ with the driving frequencies $\omega_d=2\pi\times2.16738$ MHz and $\omega_d=2\pi\times2.37940$ MHz.  $\mathrm{BW}=40$kHz}
\end{figure*}

An additional low-frequency modulation signal is added to the total voltage applied to the membrane B set of IDEs. It now writes:
$\vtot=\vdc+\vac\cos(\omega_d t)+V_p \cos(\omega_p t)$
with $V_p$ and $\omega_p\ll\omega_d$ the modulation amplitude and frequency respectively.

We reduce the set of experimental variables by locking $\vdc=2$V and $\vac=3$V while $V_p$ is used as the control parameter to explore the dynamical changes of the system. The modulation frequency is set to $\omega_p=2\pi\times7$ kHz. The driving frequency is set to the low-frequency edge of the ($-$) eigenmode bistability curve at $\omega_d=2\pi\times$2.164 MHz. The chaotic dynamics tends to disappear when the driving frequency is set apart from this position. The He-Ne laser is focused on membrane B and the modulation amplitude $V_p$ is swept from 0 to 3V. Higher values are not reached in order to preserve  the mechanical system from failure. For each value of $V_p$, we record the signal quadratures $X_{A,B}=R_{A,B} \cos(\theta_{A,B})$ and $Y_{A,B}=R_{A,B} \sin(\theta_{A,B})$ in real-time using thus accessing simultaneously phase and amplitude components. The sampling rate is 500 kHz and each trace has a length of 100ms, ensuring that several hundreds of modulation periods are recorded. In \cref{fig3:b,fig3:e}, we plot $Y_B(t)$ as a function of $X_B(t)$. It actually corresponds to a 2D projection of the whole dynamical phase space. The Poincaré section made of the local maxima of $Y_B(t)$ as a function of $V_p$ is shown in (\cref{fig3:a}). Using $Y_B (t)$ rather than $X_B (t)$ is an arbitrary choice motivated by the higher amplitude of the phase portraits along the Y axis. Additionally, each time trace is used to compute the largest Lyapunov exponent (LLE) shown below the diagram. We use the TISEAN package \cite{Tisean} routine implementing the Rosenstein algorithm \cite{ROSENSTEIN1993117}. This calculation essentially relies on the delay embedding reconstruction of the phase space in which initially close trajectories are compared over time.

For a low value of the modulation voltage injected into the normal mode ($-$), the Poincaré section in \cref{fig3:a} results in a closed single loop. In this limit-cycle oscillation regime the membranes oscillation envelopes are modulated at $\omega_p$. As the amplitude is modulated stronger, we observe two consecutive  period-doubling bifurcations at $V_p\approx1.75$V and $V_p\approx2.5$V prior to a window of chaotic dynamics for a modulation amplitude higher than $2.8$V. The presence of chaos is confirmed by the positive LLE while it is zero for limit cycle oscillations. Similar measurements are conducted driving the other normal mode ($+$). The driving frequency is now set to the low-frequency edge of the bistability at $\omega_d=2\pi\times2.379$ MHz. We construct the bifurcation diagrams still reading the motion of membrane B (\cref{fig3:d}). The phase portraits associated to this case are shown in \cref{fig3:e}. The bifurcation diagrams of eigenmode ($+$) also display a period-doubling route to chaos structure \cite{Shin1998} although the chaotic regime now occurs around $V_p\approx2$V. We observe several chaotic regions that are separated by small windows of periodic or quasiperiodic regimes as captured by the zero values of the associated LLE. An example of such regime is show in \cref{fig3:e} at value $V_p=2.6$V with a period-4 motion. Both experimental diagrams share a common dynamics but the bifurcation points significantly differ whether the eigenmode ($-$) or ($+$) is driven. This quantitative differences between the eigenmodes dynamics result from the imbalanced energy injection in the normal modes since only membrane B is driven.  
Identical measurements are performed by reading the membrane A and are shown in \cref{appen:c}. Both membranes basically settle in the same dynamical regime under a given excitation.

The bifurcation diagram for the ($-$) mechanical mode (resp. for the ($+$) mode) is numerically reproduced in \cref{fig3:c} (resp. in \cref{fig3:f}) using the Duffing-Duffing model developed previously at the driving frequency $\omega_d=2\pi\times2.16738$ MHz. (resp. at $\omega_d=2\pi\times2.37940$ MHz).
 The simulations implement an adaptative step-size RK4 method to solve the ordinary set of differential equations shown in \cref{NLsys}, including the time dependent forcing $\tilde{f}_B $ and injecting the experimentally determined parameters with
\begin{equation}\label{nonautonaumousdrive}
\tilde{f}_B (t) = f_B\Big[1+\frac{V_p}{\vdc}\cos(\omega_pt)\Big]
\end{equation}
This results in a modulated force amplitude $F_B (t)\propto\vac(\vdc+V_p \cos(\omega_p t))$. The additional terms arising from the expansion of \cref{EMforce} are either static ($V_p^2$) or slowly oscillating (at $\omega_p$ and $2\omega_p$) and do not significantly participate to the dynamics in our system. Indeed, the force associated to these terms  leads to a displacement smaller than the resonant motion by an amount given by the mechanical total quality factor $Q_m\approx660$.
The resulting time traces are analyzed with the same protocol used for our experimental data. In particular, the effect of the bandwidth demodulation is reproduced by applying an identical low-pass filter on the time traces. We use the quadrature $w_B$ which corresponds to the observable $Y_B$. The route to chaos by period doubling cascade is well captured by our model. The quantitative comparison with the experimental results yield a satisfactory agreement, which is remarkable given the experimental parameter uncertainties and the simplicity of the model. 
The period-doubling cascade bifurcation positions are corroborated with a continuation method.
Thanks to the model, it is possible to track the origin of the chaotic dynamics in the force modulation and not in the coupling: indeed uncoupled membranes also display chaos under modulation. Additional experimental and numerical analysis further show that the modulation frequency plays a role in the appearance of the chaotic regime and a large window of frequencies around the damping timescale $\Gamma_{A,B}^{-1}$ leads to chaos. However, our results show that the transfer of the chaotic dynamics from one membrane to the other is possible in spite of the detuning between the membrane resonant frequencies. 
Furthermore, the chaotic dynamics is also imprinted in the laser field intensity thanks to the integrated Fabry-Pérot cavity. This mechanical-to-optical chaos transfer is an interesting concept to exploit in chaos based technologies.

\section{NONLINEARLY COUPLED NORMAL MODES} \label{sec5}

The previous experiment shows that although the normal modes can be driven to chaotic regime using amplitude modulation, they have their own bifurcation points.  
Additionally, the eigenmodes ($-$) and ($+$) are no longer expected to be orthogonal as soon as the Duffing regime is reached. This orthogonality breaking \cite{Cadeddu2016,MercierdeLepinay2018} enables the eigenmodes to couple. In order to evidence this coupling, we take advantage of the chaotic dynamics generated by amplitude modulation. We inject energy in both normal modes by using two resonant forces ($\vac^-$ and $\vac^+$) and modulate their evolution with a common modulation.
The situation is therefore analog to driven coupled chaotic systems in which we will investigate on the synchronization properties and where the modulating signal plays the role of the drive. In this experiment, we can interestingly excite the system and measure all the dynamical variables exclusively through membrane B. The results that one would obtain by probing membrane A would be identical but with lower amplitude.

\begin{figure}[htp]
\centering
\includegraphics[scale=0.41]{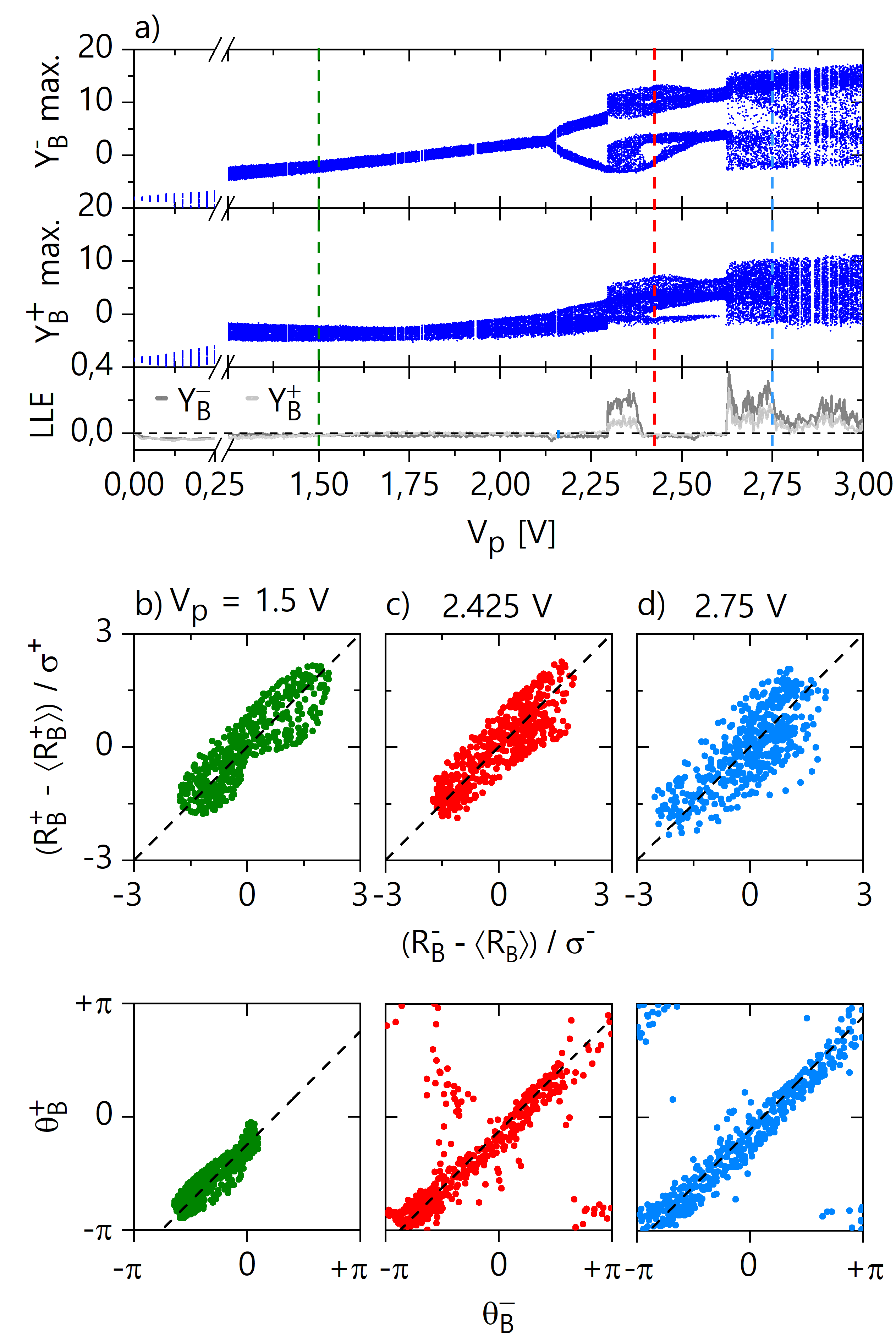}
{\phantomsubcaption\label{fig4:a}}
{\phantomsubcaption\label{fig4:b}}
{\phantomsubcaption\label{fig4:c}}
{\phantomsubcaption\label{fig4:d}}
\caption{a) experimental bifurcation diagrams and associated calculated largest Lyapunov exponent (LLE in ms$^{-1}$) built from the normal modes response quadrature $Y_B^-$ and $Y_B^+$ in units of nm$\times\sqrt{\mathrm{BW}}$ with swept parameter $V_p$. The modes are driven at $\vac^-=3.5$V and $\vac^+=0.5$V at frequencies $\omega_d^-=2\pi\times2.177$ MHz and $\omega_d^+=2\pi\times2.410$ MHz with amplitude modulation at $\omega_p=2\pi\times5$ kHz. For $V_p=1.5$V (b), $V_p=2.425$V (c) and $V_p=2.750$V (d): phase portraits showing the normal modes relative response normalized amplitudes (top) and phase (bottom) with perfect amplitude or phase synchronization references (black dashed lines). $\mathrm{BW}=40$kHz.}
\end{figure}

We adapt the previously derived equations and obtain a new system of governing equations:
\begin{equation*}
\left\{
\begin{aligned}
\ddot{x}_A + \Gamma_A \dot{x}_A + \omega_A^2x_A+\beta x_A^3-Gx_B =& 0  \\
\ddot{x}_B + \Gamma_B \dot{x}_B + \omega_B^2x_B+\beta x_B^3-Gx_A =& F_B^-\cos(\omega_d^-t) \\ 
+& F_B^+\cos(\omega_d^+t) 
\end{aligned}
\right.
\end{equation*}
\normalsize
where $F_B^-$ (resp. $F_B^+$) is the force amplitude exerted on eigenmode ($-$) with driving frequency $\omega_d^-$ (resp. eigenmode ($+$) at $\omega_d^+$).
The stationary solutions are detailed in \cref{appen:d}. These solutions highlight a nonlinear coupling between the normal modes of the form $\tilde{\beta} r^\pm r^{\mp^2}$. The non-autonomous equations resulting from the amplitude-modulation are further used for the numerical simulations by implementing the time-dependent strengths previously presented in \cref{nonautonaumousdrive}.

 The new total voltage applied on membrane B set of IDEs writes:
\begin{equation*}
\vtot = \vdc + \vac^-\cos(\omega_d^-t) + \vac^+\cos(\omega_d^+t) + V_p\cos(\omega_pt)
\end{equation*}

with $\vdc=2V$, $\vac^-=3.5V$, $\omega_d^-=2\pi\times2.177$ MHZ, $\vac^+=0.5V$, $\omega_d^+=2\pi\times2.410$ MHz and $\omega_p=2\pi\times5$ kHz. We chose $\vac^+<\vac^-$ in order to compensate the response amplitudes imbalance that results from the membranes frequency mismatch. We place the laser spot on membrane B and use two independent demodulators to simultaneously access the signal amplitude and phase at $\omega_d^-$ ($R_B^-$ and $\theta_B^-$) and at $\omega_d^+$ ($R_B^+$ and $\theta_B^+$). By sweeping the bifurcation parameter $V_p$, a new diagram is built from the local maxima of signal quadratures $Y_B^-= R_B^- \sin(\theta_B^-)$ and $Y_B^+= R_B^+\sin(\theta_B^+)$ that we record with demodulation bandwidth of 40 kHz. This  allows a reduction of the crosstalk between the channels of about -5dB. We note that the diagram branches are broader than in the single-excitation case. This is caused by the remaining crosstalks between the two demodulation channels. The qualitative comparison of the bifurcation diagrams shows a clear match of the dynamical regimes in which the normal modes ($-$) and ($+$) settle, more importantly the bifurcation points are the same. After a limit-cycle region, both display identical period-doubling route to chaos structure confirmed by the LLE computed for each diagram (see \cref{fig4:a}).

We plot the phase portraits showing the eigenmodes normalized amplitudes relative dynamics in \cref{fig4:b,fig4:c,fig4:d} (top) for three singular dynamical regimes. The normalization is meant to get rid of the unbalanced amplitudes still present despite the earlier discussed strengths adaptation. We show $(R_B^\pm-\langle R_B^\pm\rangle)/\sigma^\pm$ with $\langle R_B^\pm\rangle$ and $\sigma^\pm$ respectively the mean value and the standard deviation of $R_B^\pm(t)$ calculated over the entire time trace. The dashed black lines correspond to the synchronization regime where both normalized amplitudes are equal. Below the period-doubling bifurcation, for $V_p<V_\mathrm{PD}=2.140$V, a master-slave relation is established between the drive and each resonance so these two inescapably move in synchrony. For $V_p>V_\mathrm{PD}$ the responses are now driven in a high-order synchronization regime. Nevertheless the amplitudes are clearly correlated to each other. This is even more manifest in the chaotic regime where the amplitudes are still correlated despite their asynchrone behaviour with the drive. This regime corresponds to the chaotic synchronization of the nonlinearly coupled eigenmodes.

We now focus on the phase responses correlations shown in \cref{fig4:a,fig4:c} (bottom). In each case, we fit the data with a unit-slope line (black-dashed) corresponding to the synchronization regime $\frac{d}{dt}(\theta_B^+-\theta_B^-)=0$. These plots show a tendency of synchrone evolutions of $\theta_B^-$ and $\theta_B^+$ under force modulation for $V_p<V_\mathrm{PD}$. Contrary to the amplitudes, the synchronization of the phases is not maintained for $V_p>V_\mathrm{PD}$ as the trajectory does not only lie on a unit-slope line.

\begin{figure}[htp]
\centering
\includegraphics[scale=0.4]{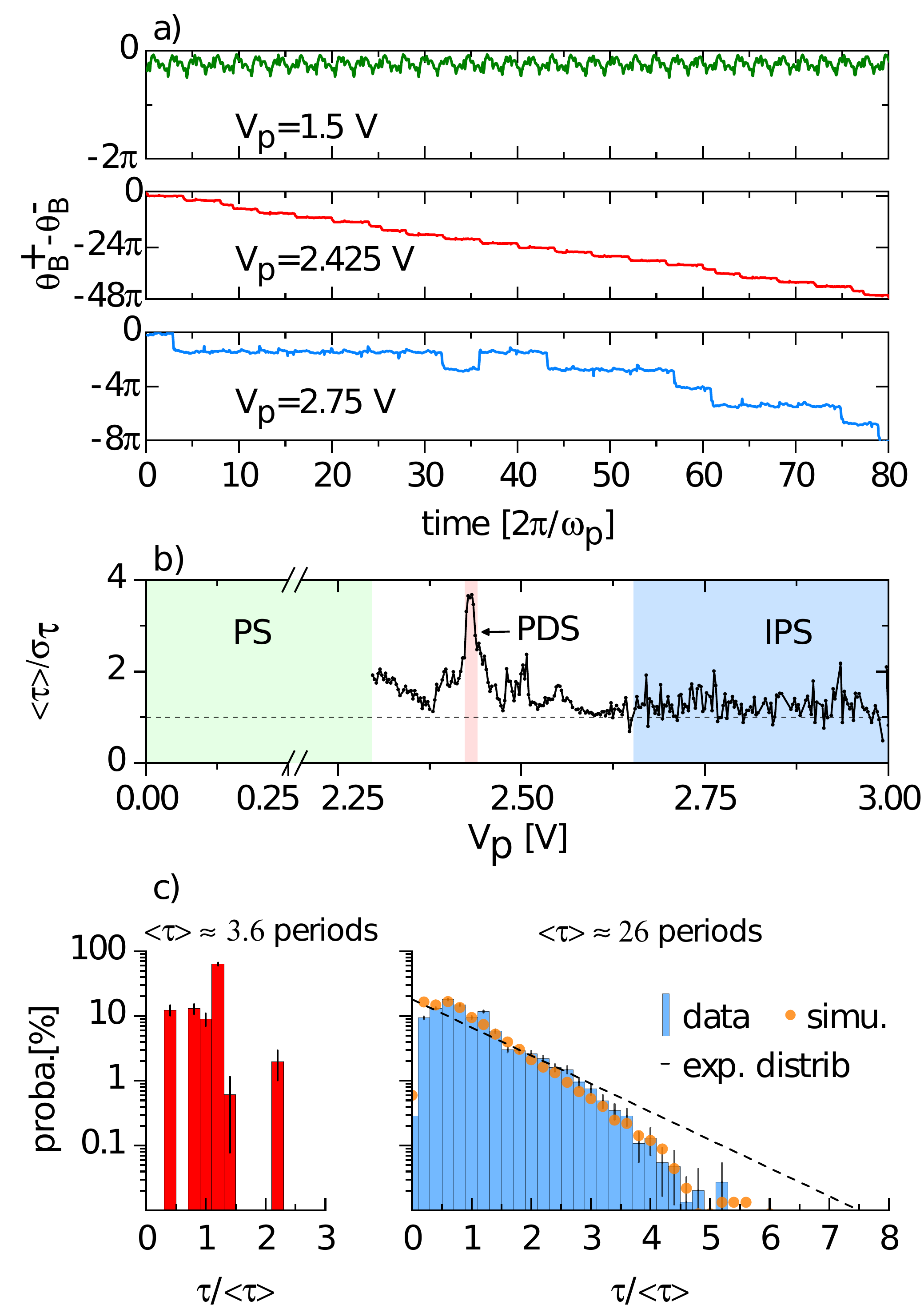}
{\phantomsubcaption\label{fig5:a}}
{\phantomsubcaption\label{fig5:b}}
{\phantomsubcaption\label{fig5:c}}

\caption{a) Measurement of the normal mode $\theta_B^+ - \theta_B^-$ over 80 modulation periods (16 \micro s). We identify phase synchronization (PS). at $V_p=1.5$V (green), phase desynchronization (PDS) at $V_p=2.425$V (red) and imperfect phase synchronization (IPS) at $V_p=2.75$V (blue). b) The scaled mean PS duration $\langle\uptau\rangle /\sigma_\uptau$ is plotted as a function of $V_p$. c) Experimental probability distributions of the PS durations within the PDS regime (red) or the IPS regime (blue). Distribution given by numerical simulation in the chaotic regime (orange dots). The exponential distribution (black dashed) is shown for comparison.}
\end{figure}

By studying the real-time dynamics of the phase difference (see \cref{fig5:a}), we find that $2\pi$ phase slips occur when $V_p>V_\mathrm{PD}$ while the resonators are phase synchronized (PS) for $V_p<V_\mathrm{PD}$ (green trace). 
When high-order synchronization is established between each mode and the drive \cite{PIK01}, phase slips resulting from phase desynchronization (PDS) can come up even if the amplitudes stay correlated. This process leads to phase slips occurring regularly (red trace) – in this situation, the phases periodically execute one more (or one less) cycle regarding the drive – or chaotically (blue trace). In the latter case, the resonator phases stay synchronized over several modulation periods and this regime is interrupted by occasional phase slips. This corresponds to the imperfect phase synchronization (IPS) scenario \cite{BOCCALETTI20021}.

The different synchronization regimes can be described through a statistical study of the durations between two successive phase slips. For a given time trace, we list all the PS durations $\uptau$ and calculate both the mean value $\langle\uptau\rangle $ as well as the standard deviation $\sigma_\uptau$. In \cref{fig5:b}, we plot the scaled mean PS duration $\langle\uptau\rangle /\sigma_\uptau$ as a function of $V_p$. No value can be estimated below the bifurcation to chaotic regime at $V_p=2.3$ V since PS is established and therefore we do not observe any phase slip in the data. The durations found to be lower than $2\pi/\omega_p$ are ignored because it can not be qualified as synchronization.

The PDS regime is identified by the regularity of the phase slips which implies that the standard deviation of the PS durations is near zero and leads to a peak in the scaled mean PS duration that can be seen around $V_p=2.425\ $V. The traces corresponding to this situation (included in the red stripe) are used to built an histogram in \cref{fig5:c} (left) showing the distribution of the PS durations probabilities with the associated 95\% confidence interval. In order to compare the statistics of $\uptau$ between the different traces, i.e. for different $V_p$, we normalize all the durations found in a given trace by the mean duration value for this trace. The probability distribution is concentrated around 1, meaning that all the phase slips have almost equal duration. The corresponding mean duration is $\langle\uptau\rangle=3.6$ modulation periods. The histogram displays a 63\% probability for the phases to synchronize during 4 modulation periods (see bar at position $\uptau/\langle\uptau\rangle=1.2$) because this PDS occurs while the systems sets in a period-4 motion dynamics. 

In the chaotic regime (blue stripe in \cref{fig5:b}) we find that the scaled mean PS duration remains constant and slightly over 1 which tends to indicate an exponential decay of the probability distribution of this quantity shown in \cref{fig5:c} (right). Note that, due to the normalization by the mean PS duration, the exponential distribution has a unique representation (black dashed curve) in this histogram. In this regime the mean PS duration is $\langle\uptau\rangle=26$ modulation periods. The probability indeed decays exponentially but we find that the probability around the mean PS duration ($\uptau/\langle\uptau\rangle=1$) is significantly higher than predicted with this distribution. Additionally the observed long PS durations occurrences are more unlikely. We conclude that the phase slips constitute a non-Poissonian process due to the deterministic chaotic dynamics and do not result from noise. The experimental histogram is reproduced using a numerical simulation realized from the new non-autonomous system of equations. We recover a similar bifurcation diagram with a robust scaling of the modulated force as shown in \cref{appen:d}. We observe $2\pi$ slips of the phase difference when the system dynamics is chaotic. From these simulations, we reproduce an histogram of the PS durations in \cref{fig5:c} integrated over a range of modulation amplitude showing a chaotic domain. We find a good agreement with the experimental distribution. Further numerical adjustments of the two driving frequencies in a restricted range evidence the possibility to achieve perfect phase synchronization in the chaotic regime. This could be of a major interest for applications based on chaos such as synchronized random number at two distinct carrier frequencies 

\section{CONCLUSION}\label{sec6}

We have analyzed the chaotic and synchronization dynamics of distinct mechanically coupled optomechanical resonators under slowly modulated near resonant drives. By setting the driving frequency at a bistability curve turning point, we show how a modulation of the driving force amplitude leads to a period-doubling cascade route to chaos. Both resonators display a chaotic dynamics, though the resonators are not synchronized. The experimentally-built bifurcation diagrams, with a direct measurement of all the dynamical variables, are numerically reproduced using a calibrated model of coupled non-identical Duffing oscillators. Because of the resonator nonlinear behavior modeled by a Duffing nonlinearity, we expect the normal modes not to be orthogonal and therefore to exchange energy. This nonlinear coupling leads to amplitude synchronization and imperfect phase synchronization as demonstrated by driving both normal modes and recording all the dynamical variables simultaneously. Within the chaotic regime, the amplitudes are locked to each other showing strong correlations. The normal modes phases dynamics are also investigated and phase synchronization, phase desynchronization and imperfect phase synchronization regimes are observed. In this last case, we perform a statistical study of the synchronization durations and the resulting non-exponential distribution is confirmed by our theoretical description attesting the deterministic nature of the dynamics. Here, we have deeply investigated the behavior of mechanically coupled optomechanical cavities. Strongly backed by simulations, we show experimental evidence of amplitude synchronization and intermittent phase synchronization of bichromatic chaotic signals opening the path toward  complex dynamics study in arrays and novel applications such as multispectral synchronized random number generation.

\begin{acknowledgments}
This work is supported by the French RENATECH network,  the European Union’s Horizon 2020 research innovation program under grant agreement No 732894 (FET Proactive HOT), the Agence Nationale de la Recherche as part of the “Investissements d’Avenir” program (Labex NanoSaclay, ANR-10-LABX-0035) with the flagship project CONDOR and the JCJC project ADOR (ANR-19-CE24-0011-01). We would like to acknowledge Marcel Clerc for fruitful discussions.
\end{acknowledgments}

\appendix 

\section{DISPLACEMENT CALIBRATION}\label{appen_calib}

A given mechanical displacement of the membrane $R_{A,B}$ converts to a measured voltage $\eta R_{A,B}$ with the conversion constant $\eta$ which can be extracted thanks to a Michelson interferometer. For that purpose, an opened-loop local oscillator is used, whose optical path difference with the sample arm is set to $\ell=\lambda/4$. Then the resonant oscillations amplitude of membrane B induced by a given driving strength, which translates to a voltage variation $\delta V$ is compared to the interferometric signal variation resulting from a small calibrated displacement $\delta \ell$ of the path difference. Doing this measurement for several excitation stage allows a confident estimation of the displacement calibration. We estimate the transduction constant to be $\eta\approx0.5$ mV.nm$^{-1}$.

\section{MECHANICAL MODES IDENTIFICATION}\label{appen:a}

\begin{figure}[htp]
\centering
\includegraphics[scale=0.4]{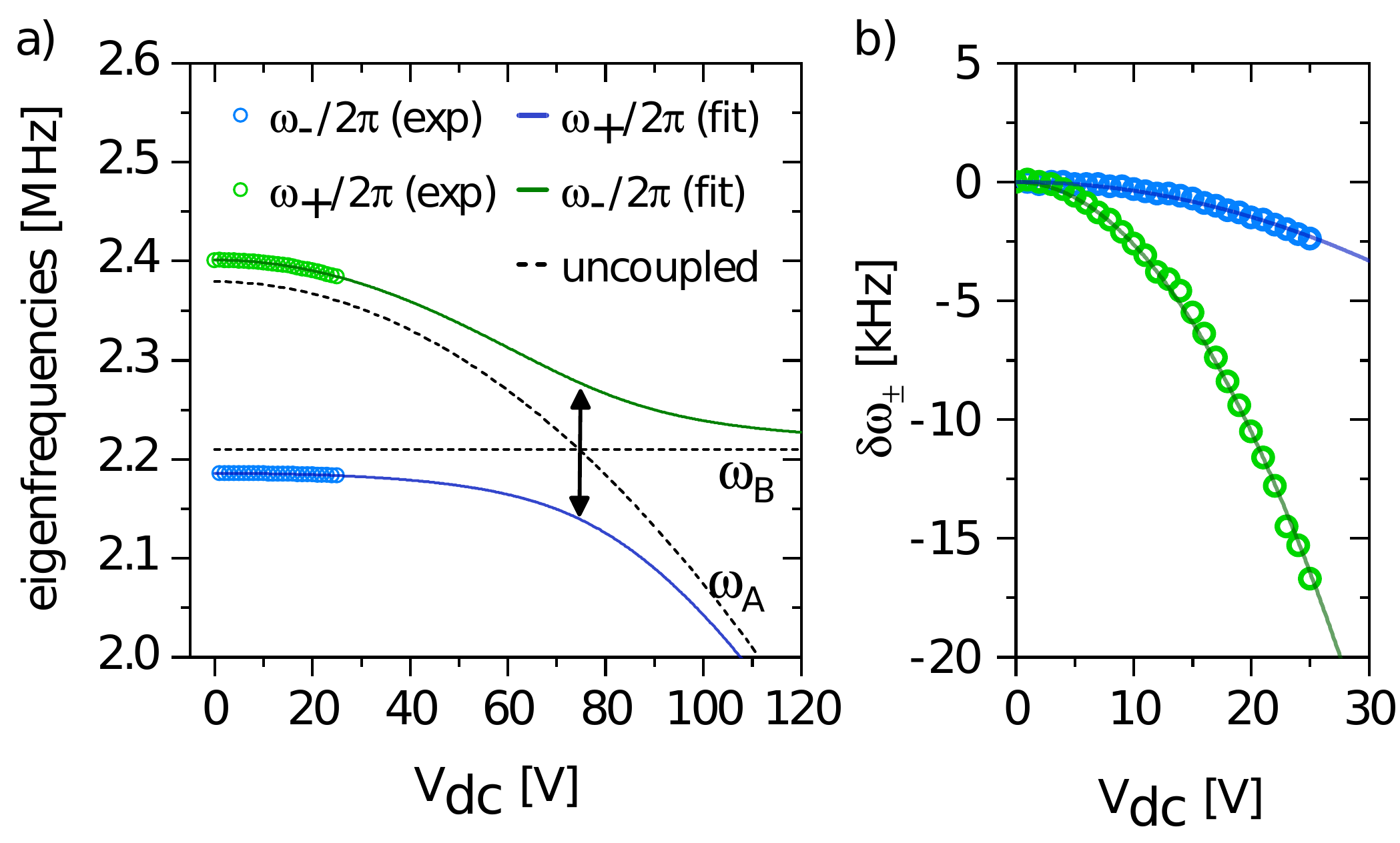}
{\phantomsubcaption\label{fig6:a}}
{\phantomsubcaption\label{fig6:b}}
\caption{a) Measurement of the eigenfrequencies under static voltage $\vdc$ applied on membrane B (symbols). The fit consider a parabolic shift of the self-coupled frequency $\omega_B$ (black dashed) that we use to solve the eigenmodes ($-$) and ($+$) (resp. blue and green lines) resulting from the previously measured coupling G. An avoided crossing is predicted at $\vdc=74$V when the natural frequencies are equal (at arrow position). b) identical experimental data presented in terms of frequencies displacement $\delta\omega_\pm=\omega_\pm(\vdc)-\omega_\pm(\vdc=0)$.}
\end{figure}

After successfully fitting membrane B response with a model of coupled harmonic oscillators (\cref{fig2:b}), we can conclude that the observed eigenmodes frequency difference is essentially caused by the natural frequency mismatch $\omega_B-\omega_A\approx2\pi\times$158 kHz. This implies that the eigenmodes ($-$) and ($+$) are respectively dominated by the motion of resonators A and B. In order the verify this conclusion, we apply a static voltage to the membrane B up to $25$V and observe how the eigenfrequencies are affected. It is well known that a static voltage acts on a micromechanical resonator as an additional residual stress \cite{RiegerAPL}. This effect can be used to tune the frequency of a oscillator and demonstrate strong-coupling between several resonators through the observation of an avoided crossing in the mechanical spectrum.
We obtain the eigenfrequencies position by sweeping the drive frequency and measuring the response spectrum. A drive amplitude such that $\vac<0.1$V is set to ensure a linear response and avoid a possible confusion with any effect of the Duffing nonlinearity. The frequency displacements we observe (\cref{fig6:a}) are not large enough to observe an avoided crossing. We compare the frequencies displacements $\delta\omega_\pm=\omega_\pm(\vdc)-\omega_\pm(\vdc=0)$  in \cref{fig6:b} and it appears clearly that the eigenmode ($+$) is mostly affected by the applied static voltage. In order to fit the data, we use the Jacobian \cite{Zanette_2018} of the linear system:
\begin{equation*}
J = \begin{pmatrix}0 & 0 & 1 & 0 \\ 0 & 0 & 0 & 1 \\ -\omega_A^2 & G & -\Gamma_A & 0 \\ G & -\omega_B^2 & 0 & -\Gamma_B\end{pmatrix}
\end{equation*}
The eigenvalues imaginary parts of $J$ correspond to the eigenfrequencies $\omega_-$ and $\omega_+$ while their real parts correspond to the eigenmodes damping rates. We input the $\vdc$-dependent self-coupled frequency $\omega_B(\vdc)=\omega_B(0)+\alpha\vdc^2$. We assume the coupling $G/\omega_B\approx2\pi\times$130 kHz and use the unchanged self-coupled frequency $\omega_B (0)$ and the coefficient $\alpha$ as the fitting parameters. We find with this second method the frequency mismatch to be $\omega_B (0)-\omega_A (0)\approx2\pi\times$168 kHz which confirms our first result.
 The resulting eigenmodes are shown with colored lines and the self-coupled frequencies with black dashed lines in \cref{fig6:a}. It appears that $\omega_A\approx\omega_-$ and $\omega_B\approx\omega_+$ in the range of $\vdc$ experimentally checked. The avoided crossing is predicted when the natural frequencies are equal, i.e. around $\vdc=74$V, which is out of reach in our experiment. Note that the assumption on the coupling value is only necessary to represent the level repulsion one would obtain with this voltage but does not affect our conclusion that the eigenmodes ($-$) and ($+$) are respectively dominated by the motions of membrane A and B.

\section{DUFFING-DUFFING MODEL}\label{appen:b}

This appendix describes the derivations of the Duffing-Duffing model shown in \cref{NLsys}. We form the master equations \cref{MasterEq} with the nonlinear potential energy:

\begin{equation*}
\left\{
\begin{aligned}
\ddot{x_A} + \Gamma_A \dot{x}_A + \omega_A^2x_A+\beta x_A^3-Gx_B =& 0  \label{CoupledDuffing} \\
\ddot{x_B} + \Gamma_B \dot{x}_B + \omega_B^2x_B+\beta x_B^3-Gx_A =& F_B\cos(\omega_dt) 
\end{aligned}
\right.
\end{equation*}

We start from the ansatz $x_A=v_A$  $\cos(\omega_d t)+ w_A \sin(\omega_d t)$ and $x_B=v_B\cos(\omega_d t)+ w_B \sin(\omega_d t)$ where the quadratures relate to the response amplitude and phase with  $r_A^2=v_A^2+w_A^2$ and $\theta_A=\atantwo(w_A,v_A)$ (and similarly for $r_B$ and $\theta_B$).
Before injecting the ansatz in the master equations, we find useful to preliminary calculate $x_A^3$ in order to reveal the off-resonant terms oscillating at $3\omega_d$ that we can neglect in the following. We also expand the expression for the derivatives $\dot{x}_A$ and $\ddot{x}_A$. We neglect the quadratures second derivatives by assuming $\ddot{v}_A$,$\ddot{w}_A \ll \omega_d^2 v_A$,$\omega_d^2w_A$. 

We now inject these preliminary results in the master equations. We use the normalized quantities $\omega_dt\rightarrow\uptau$, $(\omega_B-\omega_d)/\omega_d\rightarrow\delta$, $(\omega_B-\omega_A)/\omega_d\rightarrow\Delta\omega$, $\Gamma_{A,B}/\omega_d\rightarrow\gamma_{A,B}$, $G/\omega_d^2\rightarrow g$, $F_B/\omega_d^2\rightarrow f_B$ and $\beta/\omega_d^2\rightarrow \tilde{\beta}$.
Note that $\omega_B^2-\omega_d^2\approx\omega_d^2\delta$ and $\omega_A^2-\omega_d^2\approx2\omega_d^2(\delta-\Delta\omega)$. It writes:

\begin{figure*}
\centering
\includegraphics[scale=0.38]{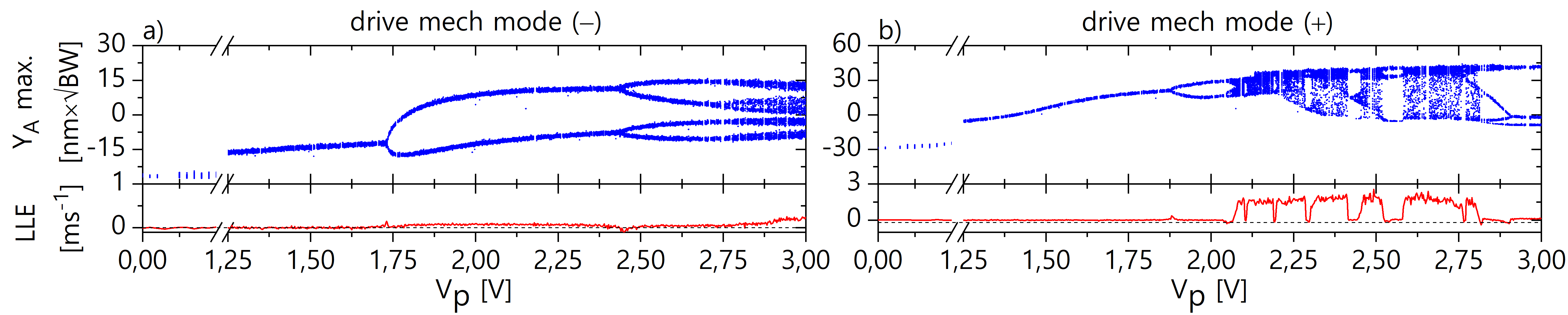}
{\phantomsubcaption\label{fig7:a}}
{\phantomsubcaption\label{fig7:b}}
\caption{Experimental bifurcation diagrams under single driving and by reading the displacement of membrane A. The measurement are performed by driving either the symmetrical (a) $\omega_d=2\pi\times2.164$ MHz) or anti-symmetrical resonance (b) $\omega_d=2\pi\times2.379$ MHz) with $\vdc=2$V, $\vac$=3V and $\omega_p=2\pi\times7$ kHz with the associated largest Lyapunov exponent (LLE).}
\end{figure*}

\begin{align*}
\dot{v}_A =& \frac{1}{2}w_A \left[ 2(\delta-\Delta\omega) + \frac{3}{4}\tilde{\beta}  (v_A^2 + w_A^2)\right]\\ -& \frac{1}{2}\gamma_A v_A - \frac{1}{2}gw_B  \tag{S.1.1}\\
\dot{w}_A =& \frac{-1}{2}v_A \left[ 2(\delta-\Delta\omega) + \frac{3}{4}\tilde{\beta}  (v_A^2 + w_A^2)\right]\\ -& \frac{1}{2}\gamma_A w_A + \frac{1}{2}gv_B  \tag{S.1.2}\\
\dot{v}_B =& \frac{1}{2}w_B \left[ 2\delta + \frac{3}{4}\tilde{\beta}  (v_B^2 + w_B^2)\right] \\ -& \frac{1}{2}\gamma_B v_B - \frac{1}{2}gw_A  \tag{S.2.1}\\
\dot{w}_B =& \frac{-1}{2}v_B \left[ 2\delta + \frac{3}{4}\tilde{\beta}  (v_B^2 + w_B^2)\right] \\ -& \frac{1}{2}\gamma_B w_B + \frac{1}{2}gv_A + \frac{1}{2}\tilde{f}_B  \tag{S.2.2}
\end{align*}

This system of equations describes the evolution of our system in terms of quadratures $v_A$, $w_A$, $v_B$ and $w_B$. It presents an interest for the numerical simulations since these quantities are homogeneous. However we can obtain the final set of equations (\cref{NLsys}) expressed in terms of the amplitudes and phases by performing the reverse transformations $v_A=r_A\cos(\theta_A )$, $w_A=r_A \sin(\theta_A)$, $v_B=r_B\cos(\theta_B)$ and $w_B=r_B\sin(\theta_B)$ and taking:
\begin{equation*}
\left\{
 \begin{aligned}
S.1.1&\times \cos(\theta_A) + S.2.1\times \sin(\theta_A)\notag\\
S.1.2&\times \cos(\theta_B) + S.2.2\times \sin(\theta_B)\notag\\
-S.1.1&\times \sin(\theta_A) + S.2.1\times \cos(\theta_A)\notag\\
-S.1.2&\times \sin(\theta_B) + S.2.2\times \cos(\theta_B)\notag
\end{aligned}
\right.
\end{equation*}

The numerical simulations are performed by integrating the ODEs with the physical quantities found in the experiments. For a given resonance, the driving frequency is adjusted near the jump-up frequency of the bistability for an optimal scaling of the bifurcation structure. The diagrams presented in \cref{fig3:c,fig3:f} are found for $\omega_d = 2\pi\times2.167380$ MHz and $\omega_d = 2\pi\times2.37940$ MHz respectively. The normalized force reads $f_B=\frac{1}{m_\mathrm{eff}\omega_d^2}\frac{dC}{dx} \vdc \vac$ with $\vdc=2$ V and $\vac=$ 3V. To account for the dephasing between the experimental driving excitation and the force, we present the simulated bifurcation diagram in \cref{fig3:f} with an offset of $80^\circ$ deg in $\theta_B$.

\section{BIFURCATION DIAGRAMS BUILT FROM THE READING OF MEMBRANE A}\label{appen:c}

The bifurcation diagrams presented in \cref{fig3:a,fig3:d} respectively show the dynamics of the normal modes ($-$) and ($+$) with parameter $V_p$. They are based on the response of membrane B. Here,we perform the same measurement by placing the laser on membrane A to read its displacement. In \cref{fig7:a} we show the bifurcation diagram made from the response of membrane A when the mode ($-$) is submitted to a modulated force by applying $\vdc$ = 2V, $\vac = 3V$ at frequency $\omega_d =2\pi\times$ 2.379 MHz and $\omega_p = 7$ kHz. The associated LLE is shown below.
Similarly in \cref{fig7:b} we show the measured diagram corresponding to the case where the normal mode ($+$) is driven with the same parameters but at resonant frequency $\omega_d =2\pi\times$ 2.164 MHz.

In both cases, the diagrams are very similar to the ones built from the reading of membrane B. In fact, the numerical simulations point out that for a given excitation regime, the membranes respond almost perfectly synchronously. This indicates that modes ($-$) and ($+$) are almost orthogonal and that measuring their response through the membrane A or B gives the same result beside the amplitude imbalance. 

Although the bifurcation diagrams are very similar whether A or B is read, a small shift in the bifurcation points positions can be observed and even a regime of periodic oscillations is present around $V_p=$2.3V in \cref{fig7:b} that is not present in \cref{fig3:d}. This is a consequence of the photothermal shift induced by the laser on the eigenfrequency dominated by the probed membrane \cite{Gao_2019}. This shift is lower than 3 kHz but leads to a significant modification of the bifurcation diagram. When driving a given normal mode, we expect the membrane responses to be perfectly correlated. The normal modes result from the strong coupling interaction between the membranes and the fact that they both are identically affected by the dynamics of a normal mode should not be understood as synchronization. This can not be confirmed without a simultaneous lecture of both membranes although it was corroborated by our numerical simulations.

\section{TWO-DRIVES MODEL AND ORTHOGONALITY BREAKING}\label{appen:d}

Driving both normal modes at the same time allows to evidence synchronization phenomena. We model the system with the same master equation but add a second resonant excitation:
\begin{equation*}
\left\{
\begin{aligned}
\ddot{x}_A + \Gamma_A \dot{x}_A + \omega_A^2(1+\tilde{\beta} x_A^2)x_A-Gx_B =& 0 \\
\ddot{x}_B + \Gamma_B \dot{x}_B + \omega_B^2(1+\tilde{\beta} x_B^2)x_B-Gx_A =& F_B^-\cos(\omega_d^-t) \\ 
+& F_B^+\cos(\omega_d^+t) 
\end{aligned}
\right.
\end{equation*}
Since the system is now expected to respond both at $\omega_d^-$ and $\omega_d^+$, we modify the ansatz:
\begin{equation*}
\begin{aligned}
x_A &= v_A^-\cos(\omega_d^-t) + w_A^- \sin(\omega_d^-t) \\
	&+ v_A^+\cos(\omega_d^+t)+w_A^+(\omega_d^+t)
\end{aligned}
\end{equation*}

The rest of the calculations is essentially the same, except for the development of the cubic terms $x_A^3$ and $x_B^3$ where the nonlinear coupling between $r_{A,B}^-$ and $r_{A,B}^+$ comes from. We neglect all off-resonant terms including the ones oscillating at $2\omega_d^\pm - \omega_d^\mp$.
Following the exact same procedure as in \cref{appen:b}, we derive a system of 8 coupled nonlinear ODEs for the normal modes ($-$) and ($+$) quadratures accessed either through the membrane A ($v_A^-$, $w_A^-$, $v_A^+$, $w_A^+$) or B ($v_B^-$, $w_B^-$, $v_B^+$, $w_B^+$). It writes:

\begin{align*}
\dot{v}_A^\pm =& \frac{1}{2}w_A^\pm \left[ 2\ep(\delta_\pm-\Delta\omega) + \frac{3}{4}\ep^2\tilde{\beta}  (r_A^{\pm^2} + 2r_A^{\mp^2})\right]\\ -& \frac{1}{2}\ep\gamma_A v_A^\pm - \frac{1}{2}\ep^2gw_B^\pm  \\
\dot{w}_A^\pm =& \frac{-1}{2}v_A^\pm \left[ 2\ep(\delta_\pm-\Delta\omega) + \frac{3}{4}\ep^2\tilde{\beta}  (r_A^{\pm^2} + 2r_A^{\mp^2}) \right]\\ -& \frac{1}{2}\ep\gamma_A w_A^\pm + \frac{1}{2}\ep^2gv_B^\pm  \\
\dot{v}_B^\pm =& \frac{1}{2}w_B^\pm \left[ 2\delta_\pm + \frac{3}{4}\ep^2\tilde{\beta}(r_B^{\pm^2} + 2r_B^{\mp^2})\right] \\ -& \frac{1}{2}\ep\gamma_B v_B^\pm - \frac{1}{2}\ep^2gw_A^\pm  \\
\dot{w}_B^\pm =& \frac{-1}{2}v_B^\pm \left[ 2\delta_\pm + \frac{3}{4}\ep^2\tilde{\beta}(r_B^{\pm^2} + 2r_B^{\mp^2})\right] \\ -& \frac{1}{2}\ep\gamma_B w_B^\pm + \frac{1}{2}\ep^2gv_A^\pm + \ep\frac{1}{2}f_B^\pm 
\end{align*}

where we use the amplitudes $r_A^{\pm^2}=v_A^{\pm^2}+w_A^{\pm^2}$ and $r_B^{\pm^2}=v_B^{\pm^2}+w_B^{\pm^2}$ for the compactness of these expressions. The parameter $\ep=\omega_d^+/\omega_d^\pm$ aims to regularize the normalization of the parameters such that the time is arbitrarily chosen to be rescaled with $\omega_d^+t$. Thus all the physical are normalized consequently to this choice: $\delta_\pm = (\omega_B-\omega_d^\pm)/\omega_d^+$, $\Delta\omega=(\omega_B-\omega_A)/\omega_d^+$, $\gamma_{A,B}=\Gamma_{A,B}/\omega_d^+$, $g=G/\omega_d^{+^2}$, $\tilde{\beta}=\beta/\omega_d^{+^2}$ and $f_B^\pm=f_B^\pm/\omega_d^+$. These two systems of 4 equations contains new terms that allow the normal modes response amplitudes $r_{A,B}^+$ and $r_{A,B}^-$ to couple. By expressing these equations in terms of amplitudes ($r_{A,B}^\pm$) and phases ($\theta_{A,B}^\pm$), the normal mode coupling takes the form $\tilde{\beta} r^\pm r^{\mp^2}$. 

\begin{figure}[htp]
\centering
\includegraphics[scale=0.38]{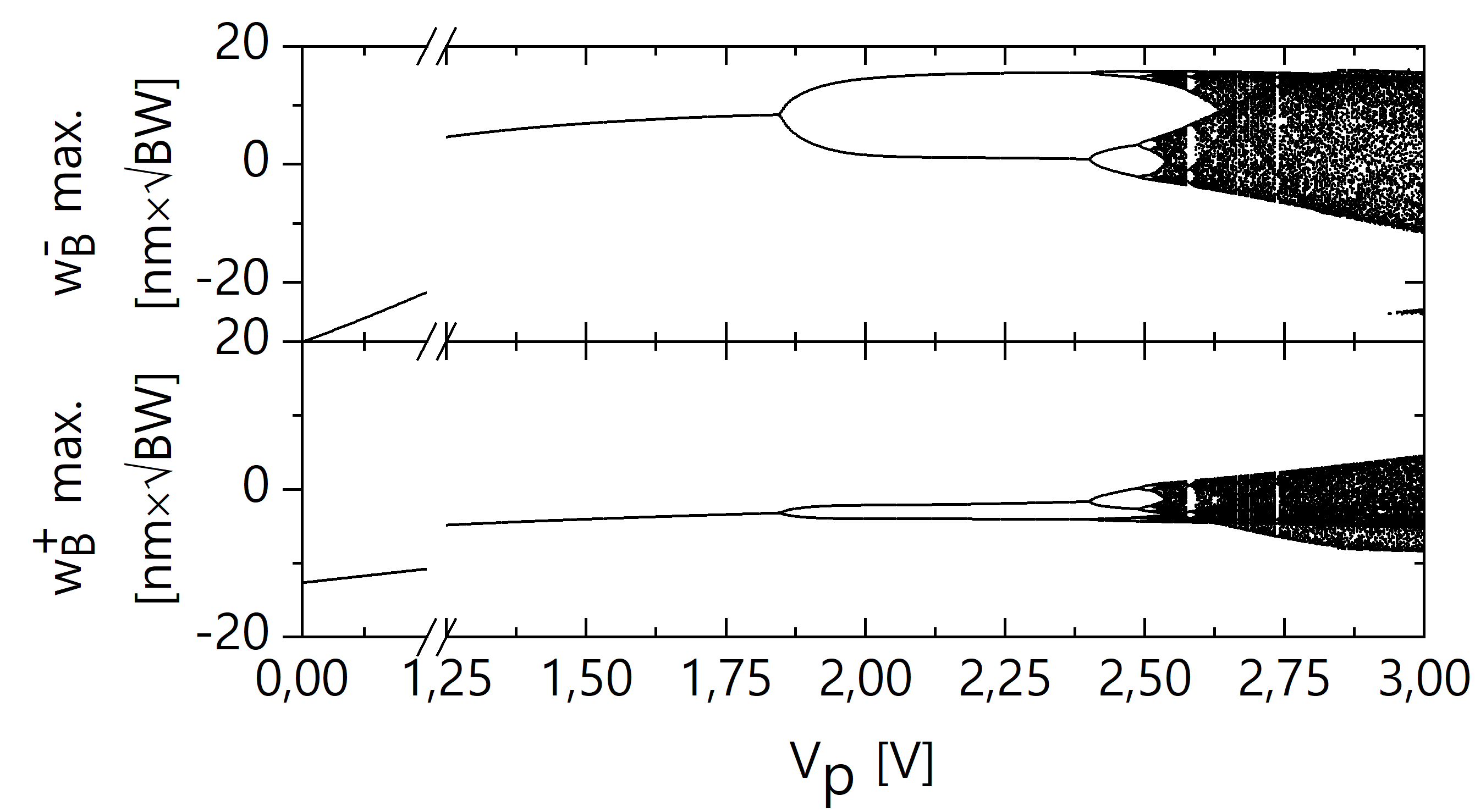}
{\phantomsubcaption\label{fig8:a}}
{\phantomsubcaption\label{fig8:b}}
\caption{Simulated bifurcation diagrams under two driving forces with parameter $V_p$. The top diagram (resp. the bottom diagram) uses the Poincaré section made from the maxima of $w_B^-$ (resp. the maxima of $w_B^+$). We use the driving frequencies $\omega_d^-=2\pi\times2.168500$ MHz, $\omega_d^+=2\pi\times2.370280$ MHz and plug the experimental parameters in the model}. This simulation should be compared with the experimental diagrams presented in \cref{fig4:a}.
\end{figure}
\vspace{3cm}

We integrate these equations using the parameters $\vdc = 2$ V, $\vac^-=3.5$ V, $\vac^+=0.5$ V, $\omega_d^-=2\pi\times2.168500$ MHz, $\omega_d^+=2\pi\times2.370280$ MHz and $\omega_p = 2\pi\times5$ kHz, V. By sweeping $V_p$, a bifurcation diagram is reproduced with the amplitude response of both normal modes using the maxima of $w_B^-$ and $w_B^+$. The phases $\theta_B^-$ and  $\theta_B^+$ are both shifted by $100^\circ$ deg thus compensating an experimental dephasing. We recover the period-doubling structure followed by chaotic intermittency that is experimentally obtained (see \cref{fig4:a}). The period doubling occurs for $V_p=1.840$ V which is slightly below the value found experimentally $V_\mathrm{PD}=2.140$ V. This can be explained by a limitation in the several experimental calibrations we have performed as well as a drift of some of the mechanical parameters with time. The amplitude of the local maxima in each simulated diagram is very similar to the corresponding experimental diagrams. Finally, the dynamical regimes in which the normal modes settle are perfectly correlated, as observed in the experiments.
We analyze a thousand chaotic traces between $V_p = 2.742$ V and $V_p = 3$ V and extract more than 22000 phase slip occurrences on which the statistical study of the phase synchronization durations presented in \cref{fig4:c} is performed.


\begin{thebibliography}{71}%
\makeatletter
\providecommand \@ifxundefined [1]{%
 \@ifx{#1\undefined}
}%
\providecommand \@ifnum [1]{%
 \ifnum #1\expandafter \@firstoftwo
 \else \expandafter \@secondoftwo
 \fi
}%
\providecommand \@ifx [1]{%
 \ifx #1\expandafter \@firstoftwo
 \else \expandafter \@secondoftwo
 \fi
}%
\providecommand \natexlab [1]{#1}%
\providecommand \enquote  [1]{``#1''}%
\providecommand \bibnamefont  [1]{#1}%
\providecommand \bibfnamefont [1]{#1}%
\providecommand \citenamefont [1]{#1}%
\providecommand \href@noop [0]{\@secondoftwo}%
\providecommand \href [0]{\begingroup \@sanitize@url \@href}%
\providecommand \@href[1]{\@@startlink{#1}\@@href}%
\providecommand \@@href[1]{\endgroup#1\@@endlink}%
\providecommand \@sanitize@url [0]{\catcode `\\12\catcode `\$12\catcode
  `\&12\catcode `\#12\catcode `\^12\catcode `\_12\catcode `\%12\relax}%
\providecommand \@@startlink[1]{}%
\providecommand \@@endlink[0]{}%
\providecommand \url  [0]{\begingroup\@sanitize@url \@url }%
\providecommand \@url [1]{\endgroup\@href {#1}{\urlprefix }}%
\providecommand \urlprefix  [0]{URL }%
\providecommand \Eprint [0]{\href }%
\providecommand \doibase [0]{https://doi.org/}%
\providecommand \selectlanguage [0]{\@gobble}%
\providecommand \bibinfo  [0]{\@secondoftwo}%
\providecommand \bibfield  [0]{\@secondoftwo}%
\providecommand \translation [1]{[#1]}%
\providecommand \BibitemOpen [0]{}%
\providecommand \bibitemStop [0]{}%
\providecommand \bibitemNoStop [0]{.\EOS\space}%
\providecommand \EOS [0]{\spacefactor3000\relax}%
\providecommand \BibitemShut  [1]{\csname bibitem#1\endcsname}%
\let\auto@bib@innerbib\@empty
\bibitem [{\citenamefont {Pikovsky}\ \emph {et~al.}(2001)\citenamefont
  {Pikovsky}, \citenamefont {Rosenblum},\ and\ \citenamefont {Kurths}}]{PIK01}%
  \BibitemOpen
  \bibfield  {author} {\bibinfo {author} {\bibfnamefont {A.}~\bibnamefont
  {Pikovsky}}, \bibinfo {author} {\bibfnamefont {M.~G.}\ \bibnamefont
  {Rosenblum}},\ and\ \bibinfo {author} {\bibfnamefont {J.}~\bibnamefont
  {Kurths}},\ }\href@noop {} {\emph {\bibinfo {title} {Synchronization, A
  Universal Concept in Nonlinear Sciences}}}\ (\bibinfo  {publisher} {Cambridge
  University Press},\ \bibinfo {address} {Cambridge},\ \bibinfo {year}
  {2001})\BibitemShut {NoStop}%
\bibitem [{\citenamefont {Huygens}(1665)}]{Huygens}%
  \BibitemOpen
  \bibfield  {author} {\bibinfo {author} {\bibfnamefont {C.}~\bibnamefont
  {Huygens}},\ }\bibfield  {title} {\bibinfo {title} {Letters to de sluse},\
  }\href@noop {} {\bibfield  {journal} {\bibinfo  {journal} {Soci{\'e}t{\'e}
  Hollandaise Des Sciences, Martinus Nijho, 1895}\ ,\ \bibinfo {pages}
  {letters; no. 1333 of 24 February 1665, no. 1335 of 26 February 1665, no.
  1345 of 6 March 1665}} (\bibinfo {year} {1665})}\BibitemShut {NoStop}%
\bibitem [{\citenamefont {{Rosenblum}}\ \emph {et~al.}(1997)\citenamefont
  {{Rosenblum}}, \citenamefont {{Pikovsky}},\ and\ \citenamefont
  {{Kurths}}}]{Rosenblum1997}%
  \BibitemOpen
  \bibfield  {author} {\bibinfo {author} {\bibfnamefont {M.~G.}\ \bibnamefont
  {{Rosenblum}}}, \bibinfo {author} {\bibfnamefont {A.~S.}\ \bibnamefont
  {{Pikovsky}}},\ and\ \bibinfo {author} {\bibfnamefont {J.}~\bibnamefont
  {{Kurths}}},\ }\bibfield  {title} {\bibinfo {title} {Phase synchronization in
  driven and coupled chaotic oscillators},\ }\href
  {https://doi.org/10.1109/81.633876} {\bibfield  {journal} {\bibinfo
  {journal} {IEEE Transactions on Circuits and Systems I: Fundamental Theory
  and Applications}\ }\textbf {\bibinfo {volume} {44}},\ \bibinfo {pages} {874}
  (\bibinfo {year} {1997})}\BibitemShut {NoStop}%
\bibitem [{\citenamefont {Pisarchik}\ and\ \citenamefont
  {Jaimes-Re{\'a}tegui}(2005)}]{Pisarchik2005}%
  \BibitemOpen
  \bibfield  {author} {\bibinfo {author} {\bibfnamefont {A.~N.}\ \bibnamefont
  {Pisarchik}}\ and\ \bibinfo {author} {\bibfnamefont {R.}~\bibnamefont
  {Jaimes-Re{\'a}tegui}},\ }\bibfield  {title} {\bibinfo {title} {Intermittent
  lag synchronization in a driven system of coupled oscillators},\ }\href
  {https://doi.org/10.1007/BF02706198} {\bibfield  {journal} {\bibinfo
  {journal} {Pramana}\ }\textbf {\bibinfo {volume} {64}},\ \bibinfo {pages}
  {503} (\bibinfo {year} {2005})}\BibitemShut {NoStop}%
\bibitem [{\citenamefont {Zhang}\ \emph {et~al.}(2012)\citenamefont {Zhang},
  \citenamefont {Wiederhecker}, \citenamefont {Manipatruni}, \citenamefont
  {Barnard}, \citenamefont {McEuen},\ and\ \citenamefont
  {Lipson}}]{ZhangPRL2012}%
  \BibitemOpen
  \bibfield  {author} {\bibinfo {author} {\bibfnamefont {M.}~\bibnamefont
  {Zhang}}, \bibinfo {author} {\bibfnamefont {G.~S.}\ \bibnamefont
  {Wiederhecker}}, \bibinfo {author} {\bibfnamefont {S.}~\bibnamefont
  {Manipatruni}}, \bibinfo {author} {\bibfnamefont {A.}~\bibnamefont
  {Barnard}}, \bibinfo {author} {\bibfnamefont {P.}~\bibnamefont {McEuen}},\
  and\ \bibinfo {author} {\bibfnamefont {M.}~\bibnamefont {Lipson}},\
  }\bibfield  {title} {\bibinfo {title} {Synchronization of micromechanical
  oscillators using light},\ }\href
  {https://doi.org/10.1103/PhysRevLett.109.233906} {\bibfield  {journal}
  {\bibinfo  {journal} {Phys. Rev. Lett.}\ }\textbf {\bibinfo {volume} {109}},\
  \bibinfo {pages} {233906} (\bibinfo {year} {2012})}\BibitemShut {NoStop}%
\bibitem [{\citenamefont {Zhang}\ \emph {et~al.}(2015)\citenamefont {Zhang},
  \citenamefont {Shah}, \citenamefont {Cardenas},\ and\ \citenamefont
  {Lipson}}]{ZhangPRL2015}%
  \BibitemOpen
  \bibfield  {author} {\bibinfo {author} {\bibfnamefont {M.}~\bibnamefont
  {Zhang}}, \bibinfo {author} {\bibfnamefont {S.}~\bibnamefont {Shah}},
  \bibinfo {author} {\bibfnamefont {J.}~\bibnamefont {Cardenas}},\ and\
  \bibinfo {author} {\bibfnamefont {M.}~\bibnamefont {Lipson}},\ }\bibfield
  {title} {\bibinfo {title} {Synchronization and phase noise reduction in
  micromechanical oscillator arrays coupled through light},\ }\href
  {https://doi.org/10.1103/PhysRevLett.115.163902} {\bibfield  {journal}
  {\bibinfo  {journal} {Phys. Rev. Lett.}\ }\textbf {\bibinfo {volume} {115}},\
  \bibinfo {pages} {163902} (\bibinfo {year} {2015})}\BibitemShut {NoStop}%
\bibitem [{\citenamefont {Kuramoto}\ and\ \citenamefont
  {Yamada}(1976)}]{Yam1976}%
  \BibitemOpen
  \bibfield  {author} {\bibinfo {author} {\bibfnamefont {Y.}~\bibnamefont
  {Kuramoto}}\ and\ \bibinfo {author} {\bibfnamefont {T.}~\bibnamefont
  {Yamada}},\ }\bibfield  {title} {\bibinfo {title} {Pattern formation in
  oscillatory chemical reactions},\ }\href {https://doi.org/10.1143/PTP.56.724}
  {\bibfield  {journal} {\bibinfo  {journal} {Progress of Theoretical Physics}\
  }\textbf {\bibinfo {volume} {56}},\ \bibinfo {pages} {724} (\bibinfo {year}
  {1976})}\BibitemShut {NoStop}%
\bibitem [{\citenamefont {Glass}(2001)}]{Glass2001}%
  \BibitemOpen
  \bibfield  {author} {\bibinfo {author} {\bibfnamefont {L.}~\bibnamefont
  {Glass}},\ }\bibfield  {title} {\bibinfo {title} {Synchronization and
  rhythmic processes in physiology},\ }\href {https://doi.org/10.1038/35065745}
  {\bibfield  {journal} {\bibinfo  {journal} {Nature}\ }\textbf {\bibinfo
  {volume} {410}},\ \bibinfo {pages} {277} (\bibinfo {year}
  {2001})}\BibitemShut {NoStop}%
\bibitem [{\citenamefont {Blasius}\ \emph {et~al.}(1999)\citenamefont
  {Blasius}, \citenamefont {Huppert},\ and\ \citenamefont
  {Stone}}]{Blasius1999}%
  \BibitemOpen
  \bibfield  {author} {\bibinfo {author} {\bibfnamefont {B.}~\bibnamefont
  {Blasius}}, \bibinfo {author} {\bibfnamefont {A.}~\bibnamefont {Huppert}},\
  and\ \bibinfo {author} {\bibfnamefont {L.}~\bibnamefont {Stone}},\ }\bibfield
   {title} {\bibinfo {title} {Complex dynamics and phase synchronization in
  spatially extended ecological systems},\ }\href
  {https://doi.org/10.1038/20676} {\bibfield  {journal} {\bibinfo  {journal}
  {Nature}\ }\textbf {\bibinfo {volume} {399}},\ \bibinfo {pages} {354}
  (\bibinfo {year} {1999})}\BibitemShut {NoStop}%
\bibitem [{\citenamefont {Volos}\ \emph {et~al.}(2012)\citenamefont {Volos},
  \citenamefont {Kyprianidis},\ and\ \citenamefont {Stouboulos}}]{Volos2012}%
  \BibitemOpen
  \bibfield  {author} {\bibinfo {author} {\bibfnamefont {C.}~\bibnamefont
  {Volos}}, \bibinfo {author} {\bibfnamefont {I.}~\bibnamefont {Kyprianidis}},\
  and\ \bibinfo {author} {\bibfnamefont {I.}~\bibnamefont {Stouboulos}},\
  }\bibfield  {title} {\bibinfo {title} {Synchronization phenomena in coupled
  nonlinear systems applied in economic cycles},\ }\href@noop {} {\bibfield
  {journal} {\bibinfo  {journal} {WSEAS Transactions on Systems}\ }\textbf
  {\bibinfo {volume} {11}},\ \bibinfo {pages} {681} (\bibinfo {year}
  {2012})}\BibitemShut {NoStop}%
\bibitem [{\citenamefont {Moussaid}\ \emph {et~al.}(2009)\citenamefont
  {Moussaid}, \citenamefont {Garnier}, \citenamefont {Theraulaz},\ and\
  \citenamefont {Helbing}}]{Moussaid2009}%
  \BibitemOpen
  \bibfield  {author} {\bibinfo {author} {\bibfnamefont {M.}~\bibnamefont
  {Moussaid}}, \bibinfo {author} {\bibfnamefont {S.}~\bibnamefont {Garnier}},
  \bibinfo {author} {\bibfnamefont {G.}~\bibnamefont {Theraulaz}},\ and\
  \bibinfo {author} {\bibfnamefont {D.}~\bibnamefont {Helbing}},\ }\bibfield
  {title} {\bibinfo {title} {Collective information processing and pattern
  formation in swarms, flocks, and crowds},\ }\href
  {https://doi.org/10.1111/j.1756-8765.2009.01028.x} {\bibfield  {journal}
  {\bibinfo  {journal} {Topics in Cognitive Science}\ }\textbf {\bibinfo
  {volume} {1}},\ \bibinfo {pages} {469} (\bibinfo {year} {2009})},\ \Eprint
  {https://arxiv.org/abs/https://onlinelibrary.wiley.com/doi/pdf/10.1111/j.1756-8765.2009.01028.x}
  {https://onlinelibrary.wiley.com/doi/pdf/10.1111/j.1756-8765.2009.01028.x}
  \BibitemShut {NoStop}%
\bibitem [{\citenamefont {Pecora}\ and\ \citenamefont
  {Caroll}(1990)}]{Pecora1990}%
  \BibitemOpen
  \bibfield  {author} {\bibinfo {author} {\bibfnamefont {L.}~\bibnamefont
  {Pecora}}\ and\ \bibinfo {author} {\bibfnamefont {T.}~\bibnamefont
  {Caroll}},\ }\bibfield  {title} {\bibinfo {title} {Synchronization in chaotic
  systems},\ }\bibfield  {journal} {\bibinfo  {journal} {Physical Review
  Letters}\ }\textbf {\bibinfo {volume} {64}},\ \href
  {https://doi.org/10.1103/PhysRevLett.64.821} {10.1103/PhysRevLett.64.821}
  (\bibinfo {year} {1990})\BibitemShut {NoStop}%
\bibitem [{\citenamefont {Duane}\ and\ \citenamefont
  {Tribbia}(2001)}]{DuanePRL}%
  \BibitemOpen
  \bibfield  {author} {\bibinfo {author} {\bibfnamefont {G.~S.}\ \bibnamefont
  {Duane}}\ and\ \bibinfo {author} {\bibfnamefont {J.~J.}\ \bibnamefont
  {Tribbia}},\ }\bibfield  {title} {\bibinfo {title} {Synchronized chaos in
  geophysical fluid dynamics},\ }\href
  {https://doi.org/10.1103/PhysRevLett.86.4298} {\bibfield  {journal} {\bibinfo
   {journal} {Phys. Rev. Lett.}\ }\textbf {\bibinfo {volume} {86}},\ \bibinfo
  {pages} {4298} (\bibinfo {year} {2001})}\BibitemShut {NoStop}%
\bibitem [{\citenamefont {Duane}\ and\ \citenamefont {Tribbia}(2004)}]{Duane}%
  \BibitemOpen
  \bibfield  {author} {\bibinfo {author} {\bibfnamefont {G.~S.}\ \bibnamefont
  {Duane}}\ and\ \bibinfo {author} {\bibfnamefont {J.~J.}\ \bibnamefont
  {Tribbia}},\ }\bibfield  {title} {\bibinfo {title} {Weak atlantic-pacific
  teleconnections as synchronized chaos},\ }\href
  {https://doi.org/10.1175/1520-0469(2004)061<2149:WATASC>2.0.CO;2} {\bibfield
  {journal} {\bibinfo  {journal} {Journal of the Atmospheric Sciences}\
  }\textbf {\bibinfo {volume} {61}},\ \bibinfo {pages} {2149} (\bibinfo {year}
  {2004})},\ \Eprint
  {https://arxiv.org/abs/https://doi.org/10.1175/1520-0469(2004)061<2149:WATASC>2.0.CO;2}
  {https://doi.org/10.1175/1520-0469(2004)061<2149:WATASC>2.0.CO;2}
  \BibitemShut {NoStop}%
\bibitem [{\citenamefont {Hiemstra}\ \emph {et~al.}(2012)\citenamefont
  {Hiemstra}, \citenamefont {Fujiwara}, \citenamefont {Selten},\ and\
  \citenamefont {Kurths}}]{Hiemstra}%
  \BibitemOpen
  \bibfield  {author} {\bibinfo {author} {\bibfnamefont {P.~H.}\ \bibnamefont
  {Hiemstra}}, \bibinfo {author} {\bibfnamefont {N.}~\bibnamefont {Fujiwara}},
  \bibinfo {author} {\bibfnamefont {F.~M.}\ \bibnamefont {Selten}},\ and\
  \bibinfo {author} {\bibfnamefont {J.}~\bibnamefont {Kurths}},\ }\bibfield
  {title} {\bibinfo {title} {Complete synchronization of chaotic atmospheric
  models by connecting only a subset of state space},\ }\href
  {https://doi.org/10.5194/npg-19-611-2012} {\bibfield  {journal} {\bibinfo
  {journal} {Nonlinear Processes in Geophysics}\ }\textbf {\bibinfo {volume}
  {19}},\ \bibinfo {pages} {611} (\bibinfo {year} {2012})}\BibitemShut
  {NoStop}%
\bibitem [{\citenamefont {Lunkeit}(2001)}]{Lunkeit}%
  \BibitemOpen
  \bibfield  {author} {\bibinfo {author} {\bibfnamefont {F.}~\bibnamefont
  {Lunkeit}},\ }\bibfield  {title} {\bibinfo {title} {Synchronization
  experiments with an atmospheric global circulation model},\ }\href
  {https://doi.org/10.1063/1.1338127} {\bibfield  {journal} {\bibinfo
  {journal} {Chaos: An Interdisciplinary Journal of Nonlinear Science}\
  }\textbf {\bibinfo {volume} {11}},\ \bibinfo {pages} {47} (\bibinfo {year}
  {2001})},\ \Eprint
  {https://arxiv.org/abs/https://aip.scitation.org/doi/pdf/10.1063/1.1338127}
  {https://aip.scitation.org/doi/pdf/10.1063/1.1338127} \BibitemShut {NoStop}%
\bibitem [{\citenamefont {Patra}\ \emph {et~al.}(2019)\citenamefont {Patra},
  \citenamefont {Altshuler},\ and\ \citenamefont {Yuzbashyan}}]{Patra}%
  \BibitemOpen
  \bibfield  {author} {\bibinfo {author} {\bibfnamefont {A.}~\bibnamefont
  {Patra}}, \bibinfo {author} {\bibfnamefont {B.~L.}\ \bibnamefont
  {Altshuler}},\ and\ \bibinfo {author} {\bibfnamefont {E.~A.}\ \bibnamefont
  {Yuzbashyan}},\ }\bibfield  {title} {\bibinfo {title} {Chaotic
  synchronization between atomic clocks},\ }\href
  {https://doi.org/10.1103/PhysRevA.100.023418} {\bibfield  {journal} {\bibinfo
   {journal} {Phys. Rev. A}\ }\textbf {\bibinfo {volume} {100}},\ \bibinfo
  {pages} {023418} (\bibinfo {year} {2019})}\BibitemShut {NoStop}%
\bibitem [{\citenamefont {Sciamanna}\ and\ \citenamefont
  {Shore}(2015)}]{shore2015}%
  \BibitemOpen
  \bibfield  {author} {\bibinfo {author} {\bibfnamefont {M.}~\bibnamefont
  {Sciamanna}}\ and\ \bibinfo {author} {\bibfnamefont {K.}~\bibnamefont
  {Shore}},\ }\bibfield  {title} {\bibinfo {title} {Physics and applications of
  laser diode chaos},\ }\href {https://doi.org/10.1038/NPHOTON.2014.326}
  {\bibfield  {journal} {\bibinfo  {journal} {Nature Photonics}\ }\textbf
  {\bibinfo {volume} {9}},\ \bibinfo {pages} {151} (\bibinfo {year}
  {2015})}\BibitemShut {NoStop}%
\bibitem [{\citenamefont {{Cuomo}}\ \emph {et~al.}(1993)\citenamefont
  {{Cuomo}}, \citenamefont {{Oppenheim}},\ and\ \citenamefont
  {{Strogatz}}}]{CuomoIEEE}%
  \BibitemOpen
  \bibfield  {author} {\bibinfo {author} {\bibfnamefont {K.~M.}\ \bibnamefont
  {{Cuomo}}}, \bibinfo {author} {\bibfnamefont {A.~V.}\ \bibnamefont
  {{Oppenheim}}},\ and\ \bibinfo {author} {\bibfnamefont {S.~H.}\ \bibnamefont
  {{Strogatz}}},\ }\bibfield  {title} {\bibinfo {title} {Synchronization of
  lorenz-based chaotic circuits with applications to communications},\ }\href
  {https://doi.org/10.1109/82.246163} {\bibfield  {journal} {\bibinfo
  {journal} {IEEE Transactions on Circuits and Systems II: Analog and Digital
  Signal Processing}\ }\textbf {\bibinfo {volume} {40}},\ \bibinfo {pages}
  {626} (\bibinfo {year} {1993})}\BibitemShut {NoStop}%
\bibitem [{\citenamefont {Argyris}\ \emph {et~al.}(2005)\citenamefont
  {Argyris}, \citenamefont {Syvridis}, \citenamefont {Larger}, \citenamefont
  {Annovazzi-Lodi}, \citenamefont {Colet}, \citenamefont {Fischer},
  \citenamefont {Garc{\'i}a-Ojalvo}, \citenamefont {Mirasso}, \citenamefont
  {Pesquera},\ and\ \citenamefont {Shore}}]{Argyris2005}%
  \BibitemOpen
  \bibfield  {author} {\bibinfo {author} {\bibfnamefont {A.}~\bibnamefont
  {Argyris}}, \bibinfo {author} {\bibfnamefont {D.}~\bibnamefont {Syvridis}},
  \bibinfo {author} {\bibfnamefont {L.}~\bibnamefont {Larger}}, \bibinfo
  {author} {\bibfnamefont {V.}~\bibnamefont {Annovazzi-Lodi}}, \bibinfo
  {author} {\bibfnamefont {P.}~\bibnamefont {Colet}}, \bibinfo {author}
  {\bibfnamefont {I.}~\bibnamefont {Fischer}}, \bibinfo {author} {\bibfnamefont
  {J.}~\bibnamefont {Garc{\'i}a-Ojalvo}}, \bibinfo {author} {\bibfnamefont
  {C.~R.}\ \bibnamefont {Mirasso}}, \bibinfo {author} {\bibfnamefont
  {L.}~\bibnamefont {Pesquera}},\ and\ \bibinfo {author} {\bibfnamefont
  {K.~A.}\ \bibnamefont {Shore}},\ }\bibfield  {title} {\bibinfo {title}
  {Chaos-based communications at high bit rates using commercial fibre-optic
  links},\ }\href {https://doi.org/10.1038/nature04275} {\bibfield  {journal}
  {\bibinfo  {journal} {Nature}\ }\textbf {\bibinfo {volume} {438}},\ \bibinfo
  {pages} {343} (\bibinfo {year} {2005})}\BibitemShut {NoStop}%
\bibitem [{\citenamefont {{Annovazzi-Lodi}}\ \emph {et~al.}(1996)\citenamefont
  {{Annovazzi-Lodi}}, \citenamefont {{Donati}},\ and\ \citenamefont
  {{Scire}}}]{Annovazzi-Lodi}%
  \BibitemOpen
  \bibfield  {author} {\bibinfo {author} {\bibfnamefont {V.}~\bibnamefont
  {{Annovazzi-Lodi}}}, \bibinfo {author} {\bibfnamefont {S.}~\bibnamefont
  {{Donati}}},\ and\ \bibinfo {author} {\bibfnamefont {A.}~\bibnamefont
  {{Scire}}},\ }\bibfield  {title} {\bibinfo {title} {Synchronization of
  chaotic injected-laser systems and its application to optical cryptography},\
  }\href {https://doi.org/10.1109/3.502371} {\bibfield  {journal} {\bibinfo
  {journal} {IEEE Journal of Quantum Electronics}\ }\textbf {\bibinfo {volume}
  {32}},\ \bibinfo {pages} {953} (\bibinfo {year} {1996})}\BibitemShut
  {NoStop}%
\bibitem [{\citenamefont {{Mirasso}}\ \emph {et~al.}(1996)\citenamefont
  {{Mirasso}}, \citenamefont {{Colet}},\ and\ \citenamefont
  {{Garcia-Fernandez}}}]{Mirasso}%
  \BibitemOpen
  \bibfield  {author} {\bibinfo {author} {\bibfnamefont {C.~R.}\ \bibnamefont
  {{Mirasso}}}, \bibinfo {author} {\bibfnamefont {P.}~\bibnamefont {{Colet}}},\
  and\ \bibinfo {author} {\bibfnamefont {P.}~\bibnamefont
  {{Garcia-Fernandez}}},\ }\bibfield  {title} {\bibinfo {title}
  {Synchronization of chaotic semiconductor lasers: application to encoded
  communications},\ }\href {https://doi.org/10.1109/68.484273} {\bibfield
  {journal} {\bibinfo  {journal} {IEEE Photonics Technology Letters}\ }\textbf
  {\bibinfo {volume} {8}},\ \bibinfo {pages} {299} (\bibinfo {year}
  {1996})}\BibitemShut {NoStop}%
\bibitem [{\citenamefont {Marinho}\ \emph {et~al.}(2005)\citenamefont
  {Marinho}, \citenamefont {Macau},\ and\ \citenamefont
  {Yoneyama}}]{MARINHO2005230}%
  \BibitemOpen
  \bibfield  {author} {\bibinfo {author} {\bibfnamefont {C.~M.}\ \bibnamefont
  {Marinho}}, \bibinfo {author} {\bibfnamefont {E.~E.}\ \bibnamefont {Macau}},\
  and\ \bibinfo {author} {\bibfnamefont {T.}~\bibnamefont {Yoneyama}},\
  }\bibfield  {title} {\bibinfo {title} {Chaos over chaos: A new approach for
  satellite communication},\ }\href
  {https://doi.org/https://doi.org/10.1016/j.actaastro.2005.03.019} {\bibfield
  {journal} {\bibinfo  {journal} {Acta Astronautica}\ }\textbf {\bibinfo
  {volume} {57}},\ \bibinfo {pages} {230 } (\bibinfo {year} {2005})},\ \bibinfo
  {note} {infinite Possibilities Global Realities, Selected Proceedings of the
  55th International Astronautical Federation Congress, Vancouver, Canada, 4-8
  October 2004}\BibitemShut {NoStop}%
\bibitem [{\citenamefont {{Hsu}}\ \emph {et~al.}(1996)\citenamefont {{Hsu}},
  \citenamefont {{Gobovic}}, \citenamefont {{Zaghloul}},\ and\ \citenamefont
  {{Szu}}}]{Hsu}%
  \BibitemOpen
  \bibfield  {author} {\bibinfo {author} {\bibfnamefont {C.~C.}\ \bibnamefont
  {{Hsu}}}, \bibinfo {author} {\bibfnamefont {D.}~\bibnamefont {{Gobovic}}},
  \bibinfo {author} {\bibfnamefont {M.~E.}\ \bibnamefont {{Zaghloul}}},\ and\
  \bibinfo {author} {\bibfnamefont {H.~H.}\ \bibnamefont {{Szu}}},\ }\bibfield
  {title} {\bibinfo {title} {Chaotic neuron models and their vlsi circuit
  implementations},\ }\href {https://doi.org/10.1109/72.548163} {\bibfield
  {journal} {\bibinfo  {journal} {IEEE Transactions on Neural Networks}\
  }\textbf {\bibinfo {volume} {7}},\ \bibinfo {pages} {1339} (\bibinfo {year}
  {1996})}\BibitemShut {NoStop}%
\bibitem [{\citenamefont {{Milanovic}}\ and\ \citenamefont
  {{Zaghloul}}(1996)}]{Milanovic}%
  \BibitemOpen
  \bibfield  {author} {\bibinfo {author} {\bibfnamefont {V.}~\bibnamefont
  {{Milanovic}}}\ and\ \bibinfo {author} {\bibfnamefont {M.~E.}\ \bibnamefont
  {{Zaghloul}}},\ }\bibfield  {title} {\bibinfo {title} {Synchronization of
  chaotic neural networks for secure communications},\ }in\ \href
  {https://doi.org/10.1109/ISCAS.1996.541472} {\emph {\bibinfo {booktitle}
  {1996 IEEE International Symposium on Circuits and Systems. Circuits and
  Systems Connecting the World. ISCAS 96}}},\ Vol.~\bibinfo {volume} {3}\
  (\bibinfo {year} {1996})\ pp.\ \bibinfo {pages} {28--31 vol.3}\BibitemShut
  {NoStop}%
\bibitem [{\citenamefont {Fiderer}\ and\ \citenamefont
  {Braun}(2018)}]{Fiderer2018}%
  \BibitemOpen
  \bibfield  {author} {\bibinfo {author} {\bibfnamefont {L.~J.}\ \bibnamefont
  {Fiderer}}\ and\ \bibinfo {author} {\bibfnamefont {D.}~\bibnamefont
  {Braun}},\ }\bibfield  {title} {\bibinfo {title} {Quantum metrology with
  quantum-chaotic sensors},\ }\href
  {https://doi.org/10.1038/s41467-018-03623-z} {\bibfield  {journal} {\bibinfo
  {journal} {Nature Communications}\ }\textbf {\bibinfo {volume} {9}},\
  \bibinfo {pages} {1351} (\bibinfo {year} {2018})}\BibitemShut {NoStop}%
\bibitem [{\citenamefont {Bakemeier}\ \emph {et~al.}(2015)\citenamefont
  {Bakemeier}, \citenamefont {Alvermann},\ and\ \citenamefont
  {Fehske}}]{PhysRevLett.114.013601}%
  \BibitemOpen
  \bibfield  {author} {\bibinfo {author} {\bibfnamefont {L.}~\bibnamefont
  {Bakemeier}}, \bibinfo {author} {\bibfnamefont {A.}~\bibnamefont
  {Alvermann}},\ and\ \bibinfo {author} {\bibfnamefont {H.}~\bibnamefont
  {Fehske}},\ }\bibfield  {title} {\bibinfo {title} {Route to chaos in
  optomechanics},\ }\href {https://doi.org/10.1103/PhysRevLett.114.013601}
  {\bibfield  {journal} {\bibinfo  {journal} {Phys. Rev. Lett.}\ }\textbf
  {\bibinfo {volume} {114}},\ \bibinfo {pages} {013601} (\bibinfo {year}
  {2015})}\BibitemShut {NoStop}%
\bibitem [{\citenamefont {Park}\ \emph {et~al.}(1999)\citenamefont {Park},
  \citenamefont {Zaks},\ and\ \citenamefont {Kurths}}]{ParkPRE}%
  \BibitemOpen
  \bibfield  {author} {\bibinfo {author} {\bibfnamefont {E.-H.}\ \bibnamefont
  {Park}}, \bibinfo {author} {\bibfnamefont {M.~A.}\ \bibnamefont {Zaks}},\
  and\ \bibinfo {author} {\bibfnamefont {J.}~\bibnamefont {Kurths}},\
  }\bibfield  {title} {\bibinfo {title} {Phase synchronization in the forced
  lorenz system},\ }\href {https://doi.org/10.1103/PhysRevE.60.6627} {\bibfield
   {journal} {\bibinfo  {journal} {Phys. Rev. E}\ }\textbf {\bibinfo {volume}
  {60}},\ \bibinfo {pages} {6627} (\bibinfo {year} {1999})}\BibitemShut
  {NoStop}%
\bibitem [{\citenamefont {Lifshitz}\ and\ \citenamefont
  {Cross}(2003)}]{Cross2003}%
  \BibitemOpen
  \bibfield  {author} {\bibinfo {author} {\bibfnamefont {R.}~\bibnamefont
  {Lifshitz}}\ and\ \bibinfo {author} {\bibfnamefont {M.}~\bibnamefont
  {Cross}},\ }\bibfield  {title} {\bibinfo {title} {Response of parametrically
  driven nonlinear coupled oscillators with application to micromechanical and
  nanomechanical resonator arrays},\ }\bibfield  {journal} {\bibinfo  {journal}
  {Physical Review B}\ }\textbf {\bibinfo {volume} {67}},\ \href
  {https://doi.org/10.1103/PhysRevB.67.134302} {10.1103/PhysRevB.67.134302}
  (\bibinfo {year} {2003})\BibitemShut {NoStop}%
\bibitem [{\citenamefont {Blackburn}\ \emph {et~al.}(2000)\citenamefont
  {Blackburn}, \citenamefont {Baker},\ and\ \citenamefont
  {Smith}}]{BlackburnPRB}%
  \BibitemOpen
  \bibfield  {author} {\bibinfo {author} {\bibfnamefont {J.~A.}\ \bibnamefont
  {Blackburn}}, \bibinfo {author} {\bibfnamefont {G.~L.}\ \bibnamefont
  {Baker}},\ and\ \bibinfo {author} {\bibfnamefont {H.~J.~T.}\ \bibnamefont
  {Smith}},\ }\bibfield  {title} {\bibinfo {title} {Intermittent
  synchronization of resistively coupled chaotic josephson junctions},\ }\href
  {https://doi.org/10.1103/PhysRevB.62.5931} {\bibfield  {journal} {\bibinfo
  {journal} {Phys. Rev. B}\ }\textbf {\bibinfo {volume} {62}},\ \bibinfo
  {pages} {5931} (\bibinfo {year} {2000})}\BibitemShut {NoStop}%
\bibitem [{\citenamefont {Wu}\ and\ \citenamefont {Chua}(1995)}]{Wu1995}%
  \BibitemOpen
  \bibfield  {author} {\bibinfo {author} {\bibfnamefont {C.}~\bibnamefont
  {Wu}}\ and\ \bibinfo {author} {\bibfnamefont {L.}~\bibnamefont {Chua}},\
  }\bibfield  {title} {\bibinfo {title} {Synchronization in an array of
  linearly coupled dynamical systems},\ }\bibfield  {journal} {\bibinfo
  {journal} {IEEE Transactions on Circuits and Systems I: Fundamental Theory
  and Applications}\ }\textbf {\bibinfo {volume} {42}},\ \href
  {https://doi.org/10.1109/81.404047} {10.1109/81.404047} (\bibinfo {year}
  {1995})\BibitemShut {NoStop}%
\bibitem [{\citenamefont {Shuai}\ and\ \citenamefont
  {Durand}(1999)}]{SHUAI1999289}%
  \BibitemOpen
  \bibfield  {author} {\bibinfo {author} {\bibfnamefont {J.-W.}\ \bibnamefont
  {Shuai}}\ and\ \bibinfo {author} {\bibfnamefont {D.~M.}\ \bibnamefont
  {Durand}},\ }\bibfield  {title} {\bibinfo {title} {Phase synchronization in
  two coupled chaotic neurons},\ }\href
  {https://doi.org/https://doi.org/10.1016/S0375-9601(99)00816-6} {\bibfield
  {journal} {\bibinfo  {journal} {Physics Letters A}\ }\textbf {\bibinfo
  {volume} {264}},\ \bibinfo {pages} {289 } (\bibinfo {year}
  {1999})}\BibitemShut {NoStop}%
\bibitem [{\citenamefont {Clerc}\ \emph {et~al.}(2018)\citenamefont {Clerc},
  \citenamefont {Coulibaly}, \citenamefont {Ferr{\'e}},\ and\ \citenamefont
  {Rojas}}]{Clerc2018}%
  \BibitemOpen
  \bibfield  {author} {\bibinfo {author} {\bibfnamefont {M.~G.}\ \bibnamefont
  {Clerc}}, \bibinfo {author} {\bibfnamefont {S.}~\bibnamefont {Coulibaly}},
  \bibinfo {author} {\bibfnamefont {M.~A.}\ \bibnamefont {Ferr{\'e}}},\ and\
  \bibinfo {author} {\bibfnamefont {R.~G.}\ \bibnamefont {Rojas}},\ }\bibfield
  {title} {\bibinfo {title} {Chimera states in a duffing oscillators chain
  coupled to nearest neighbors},\ }\href {https://doi.org/10.1063/1.5025038}
  {\bibfield  {journal} {\bibinfo  {journal} {Chaos: An Interdisciplinary
  Journal of Nonlinear Science}\ }\textbf {\bibinfo {volume} {28}},\ \bibinfo
  {pages} {083126} (\bibinfo {year} {2018})},\ \Eprint
  {https://arxiv.org/abs/https://doi.org/10.1063/1.5025038}
  {https://doi.org/10.1063/1.5025038} \BibitemShut {NoStop}%
\bibitem [{\citenamefont {Martens}\ \emph {et~al.}(2013)\citenamefont
  {Martens}, \citenamefont {Thutupalli}, \citenamefont {Fourri{\`e}re},\ and\
  \citenamefont {Hallatschek}}]{Martens10563}%
  \BibitemOpen
  \bibfield  {author} {\bibinfo {author} {\bibfnamefont {E.~A.}\ \bibnamefont
  {Martens}}, \bibinfo {author} {\bibfnamefont {S.}~\bibnamefont {Thutupalli}},
  \bibinfo {author} {\bibfnamefont {A.}~\bibnamefont {Fourri{\`e}re}},\ and\
  \bibinfo {author} {\bibfnamefont {O.}~\bibnamefont {Hallatschek}},\
  }\bibfield  {title} {\bibinfo {title} {Chimera states in mechanical
  oscillator networks},\ }\href {https://doi.org/10.1073/pnas.1302880110}
  {\bibfield  {journal} {\bibinfo  {journal} {Proceedings of the National
  Academy of Sciences}\ }\textbf {\bibinfo {volume} {110}},\ \bibinfo {pages}
  {10563} (\bibinfo {year} {2013})},\ \Eprint
  {https://arxiv.org/abs/https://www.pnas.org/content/110/26/10563.full.pdf}
  {https://www.pnas.org/content/110/26/10563.full.pdf} \BibitemShut {NoStop}%
\bibitem [{\citenamefont {Pelka}\ \emph {et~al.}(2019)\citenamefont {Pelka},
  \citenamefont {Peano},\ and\ \citenamefont {Xuereb}}]{pelka2019chimera}%
  \BibitemOpen
  \bibfield  {author} {\bibinfo {author} {\bibfnamefont {K.}~\bibnamefont
  {Pelka}}, \bibinfo {author} {\bibfnamefont {V.}~\bibnamefont {Peano}},\ and\
  \bibinfo {author} {\bibfnamefont {A.}~\bibnamefont {Xuereb}},\ }\href@noop {}
  {\bibinfo {title} {Chimera states in small disordered optomechanical arrays}}
  (\bibinfo {year} {2019}),\ \Eprint {https://arxiv.org/abs/1905.01115}
  {arXiv:1905.01115 [quant-ph]} \BibitemShut {NoStop}%
\bibitem [{\citenamefont {Lauter}\ \emph {et~al.}(2015)\citenamefont {Lauter},
  \citenamefont {Brendel}, \citenamefont {Habraken},\ and\ \citenamefont
  {Marquardt}}]{Marquardt2015}%
  \BibitemOpen
  \bibfield  {author} {\bibinfo {author} {\bibfnamefont {R.}~\bibnamefont
  {Lauter}}, \bibinfo {author} {\bibfnamefont {C.}~\bibnamefont {Brendel}},
  \bibinfo {author} {\bibfnamefont {S.~J.~M.}\ \bibnamefont {Habraken}},\ and\
  \bibinfo {author} {\bibfnamefont {F.}~\bibnamefont {Marquardt}},\ }\bibfield
  {title} {\bibinfo {title} {Pattern phase diagram for two-dimensional arrays
  of coupled limit-cycle oscillators},\ }\href
  {https://doi.org/10.1103/PhysRevE.92.012902} {\bibfield  {journal} {\bibinfo
  {journal} {Phys. Rev. E}\ }\textbf {\bibinfo {volume} {92}},\ \bibinfo
  {pages} {012902} (\bibinfo {year} {2015})}\BibitemShut {NoStop}%
\bibitem [{\citenamefont {Akhmedov}\ and\ \citenamefont
  {Mirizzi}(2016)}]{Mirizzi2016}%
  \BibitemOpen
  \bibfield  {author} {\bibinfo {author} {\bibfnamefont {E.}~\bibnamefont
  {Akhmedov}}\ and\ \bibinfo {author} {\bibfnamefont {A.}~\bibnamefont
  {Mirizzi}},\ }\bibfield  {title} {\bibinfo {title} {Another look at
  synchronized neutrino oscillations},\ }\href
  {https://doi.org/10.1016/j.nuclphysb.2016.02.011} {\bibfield  {journal}
  {\bibinfo  {journal} {Nuclear Physics B}\ }\textbf {\bibinfo {volume}
  {908}},\ \bibinfo {pages} {382} (\bibinfo {year} {2016})}\BibitemShut
  {NoStop}%
\bibitem [{\citenamefont {Midolo}\ \emph {et~al.}(2018)\citenamefont {Midolo},
  \citenamefont {Schliesser},\ and\ \citenamefont {Fiore}}]{Midolo2018}%
  \BibitemOpen
  \bibfield  {author} {\bibinfo {author} {\bibfnamefont {L.}~\bibnamefont
  {Midolo}}, \bibinfo {author} {\bibfnamefont {A.}~\bibnamefont {Schliesser}},\
  and\ \bibinfo {author} {\bibfnamefont {A.}~\bibnamefont {Fiore}},\ }\bibfield
   {title} {\bibinfo {title} {Nano-opto-electro-mechanical systems},\ }\href
  {https://doi.org/10.1038/s41565-017-0039-1} {\bibfield  {journal} {\bibinfo
  {journal} {Nature Nanotechnology}\ }\textbf {\bibinfo {volume} {13}},\
  \bibinfo {pages} {11} (\bibinfo {year} {2018})}\BibitemShut {NoStop}%
\bibitem [{\citenamefont {Gao}\ \emph {et~al.}(2019)\citenamefont {Gao},
  \citenamefont {Luo}, \citenamefont {Yan}, \citenamefont {Zhang},\ and\
  \citenamefont {Liu}}]{Gao_2019}%
  \BibitemOpen
  \bibfield  {author} {\bibinfo {author} {\bibfnamefont {N.}~\bibnamefont
  {Gao}}, \bibinfo {author} {\bibfnamefont {W.}~\bibnamefont {Luo}}, \bibinfo
  {author} {\bibfnamefont {W.}~\bibnamefont {Yan}}, \bibinfo {author}
  {\bibfnamefont {D.}~\bibnamefont {Zhang}},\ and\ \bibinfo {author}
  {\bibfnamefont {D.}~\bibnamefont {Liu}},\ }\bibfield  {title} {\bibinfo
  {title} {Continuously tuning the resonant characteristics of microcantilevers
  by a laser induced photothermal effect},\ }\href
  {https://doi.org/10.1088/1361-6463/ab2608} {\bibfield  {journal} {\bibinfo
  {journal} {Journal of Physics D: Applied Physics}\ }\textbf {\bibinfo
  {volume} {52}},\ \bibinfo {pages} {385402} (\bibinfo {year}
  {2019})}\BibitemShut {NoStop}%
\bibitem [{\citenamefont {Unterreithmeier}\ \emph {et~al.}(2009)\citenamefont
  {Unterreithmeier}, \citenamefont {Weig},\ and\ \citenamefont
  {Kotthaus}}]{Unterreithmeier2009}%
  \BibitemOpen
  \bibfield  {author} {\bibinfo {author} {\bibfnamefont {Q.~P.}\ \bibnamefont
  {Unterreithmeier}}, \bibinfo {author} {\bibfnamefont {E.~M.}\ \bibnamefont
  {Weig}},\ and\ \bibinfo {author} {\bibfnamefont {J.~P.}\ \bibnamefont
  {Kotthaus}},\ }\bibfield  {title} {\bibinfo {title} {Universal transduction
  scheme for nanomechanical systems based on dielectric forces},\ }\href
  {https://doi.org/10.1038/nature07932} {\bibfield  {journal} {\bibinfo
  {journal} {Nature}\ }\textbf {\bibinfo {volume} {458}},\ \bibinfo {pages}
  {1001} (\bibinfo {year} {2009})}\BibitemShut {NoStop}%
\bibitem [{\citenamefont {Eichler}\ \emph {et~al.}(2012)\citenamefont
  {Eichler}, \citenamefont {del \'Alamo~Ruiz}, \citenamefont {Plaza},\ and\
  \citenamefont {Bachtold}}]{Eichler2012}%
  \BibitemOpen
  \bibfield  {author} {\bibinfo {author} {\bibfnamefont {A.}~\bibnamefont
  {Eichler}}, \bibinfo {author} {\bibfnamefont {M.}~\bibnamefont {del
  \'Alamo~Ruiz}}, \bibinfo {author} {\bibfnamefont {J.~A.}\ \bibnamefont
  {Plaza}},\ and\ \bibinfo {author} {\bibfnamefont {A.}~\bibnamefont
  {Bachtold}},\ }\bibfield  {title} {\bibinfo {title} {Strong coupling between
  mechanical modes in a nanotube resonator},\ }\href
  {https://doi.org/10.1103/PhysRevLett.109.025503} {\bibfield  {journal}
  {\bibinfo  {journal} {Phys. Rev. Lett.}\ }\textbf {\bibinfo {volume} {109}},\
  \bibinfo {pages} {025503} (\bibinfo {year} {2012})}\BibitemShut {NoStop}%
\bibitem [{\citenamefont {Gajo}\ \emph {et~al.}(2017)\citenamefont {Gajo},
  \citenamefont {Sch{\"u}z},\ and\ \citenamefont {Weig}}]{Gajo2017}%
  \BibitemOpen
  \bibfield  {author} {\bibinfo {author} {\bibfnamefont {K.}~\bibnamefont
  {Gajo}}, \bibinfo {author} {\bibfnamefont {S.}~\bibnamefont {Sch{\"u}z}},\
  and\ \bibinfo {author} {\bibfnamefont {E.~M.}\ \bibnamefont {Weig}},\
  }\bibfield  {title} {\bibinfo {title} {Strong 4-mode coupling of
  nanomechanical string resonators},\ }\href
  {https://doi.org/10.1063/1.4995230} {\bibfield  {journal} {\bibinfo
  {journal} {Applied Physics Letters}\ }\textbf {\bibinfo {volume} {111}},\
  \bibinfo {pages} {133109} (\bibinfo {year} {2017})},\ \Eprint
  {https://arxiv.org/abs/https://doi.org/10.1063/1.4995230}
  {https://doi.org/10.1063/1.4995230} \BibitemShut {NoStop}%
\bibitem [{\citenamefont {Okamoto}\ \emph {et~al.}(2013)\citenamefont
  {Okamoto}, \citenamefont {Gourgout}, \citenamefont {Chang}, \citenamefont
  {Onomitsu}, \citenamefont {Mahboob}, \citenamefont {Chang},\ and\
  \citenamefont {Yamaguchi}}]{Okamoto2013}%
  \BibitemOpen
  \bibfield  {author} {\bibinfo {author} {\bibfnamefont {H.}~\bibnamefont
  {Okamoto}}, \bibinfo {author} {\bibfnamefont {A.}~\bibnamefont {Gourgout}},
  \bibinfo {author} {\bibfnamefont {C.-Y.}\ \bibnamefont {Chang}}, \bibinfo
  {author} {\bibfnamefont {K.}~\bibnamefont {Onomitsu}}, \bibinfo {author}
  {\bibfnamefont {I.}~\bibnamefont {Mahboob}}, \bibinfo {author} {\bibfnamefont
  {E.~Y.}\ \bibnamefont {Chang}},\ and\ \bibinfo {author} {\bibfnamefont
  {H.}~\bibnamefont {Yamaguchi}},\ }\bibfield  {title} {\bibinfo {title}
  {Coherent phonon manipulation in coupled mechanical resonators},\ }\href
  {https://doi.org/10.1038/nphys2665} {\bibfield  {journal} {\bibinfo
  {journal} {Nature Physics}\ }\textbf {\bibinfo {volume} {9}},\ \bibinfo
  {pages} {480} (\bibinfo {year} {2013})}\BibitemShut {NoStop}%
\bibitem [{\citenamefont {Chowdhury}\ \emph {et~al.}(2017)\citenamefont
  {Chowdhury}, \citenamefont {Barbay}, \citenamefont {Clerc}, \citenamefont
  {Robert-Philip},\ and\ \citenamefont {Braive}}]{chowdhuryPRL}%
  \BibitemOpen
  \bibfield  {author} {\bibinfo {author} {\bibfnamefont {A.}~\bibnamefont
  {Chowdhury}}, \bibinfo {author} {\bibfnamefont {S.}~\bibnamefont {Barbay}},
  \bibinfo {author} {\bibfnamefont {M.~G.}\ \bibnamefont {Clerc}}, \bibinfo
  {author} {\bibfnamefont {I.}~\bibnamefont {Robert-Philip}},\ and\ \bibinfo
  {author} {\bibfnamefont {R.}~\bibnamefont {Braive}},\ }\bibfield  {title}
  {\bibinfo {title} {Phase stochastic resonance in a forced
  nanoelectromechanical membrane},\ }\href
  {https://doi.org/10.1103/PhysRevLett.119.234101} {\bibfield  {journal}
  {\bibinfo  {journal} {Phys. Rev. Lett.}\ }\textbf {\bibinfo {volume} {119}},\
  \bibinfo {pages} {234101} (\bibinfo {year} {2017})}\BibitemShut {NoStop}%
\bibitem [{\citenamefont {Shim}\ \emph {et~al.}(2007)\citenamefont {Shim},
  \citenamefont {Imboden},\ and\ \citenamefont {Mohanty}}]{Shim95}%
  \BibitemOpen
  \bibfield  {author} {\bibinfo {author} {\bibfnamefont {S.-B.}\ \bibnamefont
  {Shim}}, \bibinfo {author} {\bibfnamefont {M.}~\bibnamefont {Imboden}},\ and\
  \bibinfo {author} {\bibfnamefont {P.}~\bibnamefont {Mohanty}},\ }\bibfield
  {title} {\bibinfo {title} {Synchronized oscillation in coupled nanomechanical
  oscillators},\ }\href {https://doi.org/10.1126/science.1137307} {\bibfield
  {journal} {\bibinfo  {journal} {Science}\ }\textbf {\bibinfo {volume}
  {316}},\ \bibinfo {pages} {95} (\bibinfo {year} {2007})},\ \Eprint
  {https://arxiv.org/abs/https://science.sciencemag.org/content/316/5821/95.full.pdf}
  {https://science.sciencemag.org/content/316/5821/95.full.pdf} \BibitemShut
  {NoStop}%
\bibitem [{\citenamefont {Pu}\ \emph {et~al.}(2018)\citenamefont {Pu},
  \citenamefont {Wei}, \citenamefont {Xu}, \citenamefont {Jiang},\ and\
  \citenamefont {Huan}}]{DongAPL}%
  \BibitemOpen
  \bibfield  {author} {\bibinfo {author} {\bibfnamefont {D.}~\bibnamefont
  {Pu}}, \bibinfo {author} {\bibfnamefont {X.}~\bibnamefont {Wei}}, \bibinfo
  {author} {\bibfnamefont {L.}~\bibnamefont {Xu}}, \bibinfo {author}
  {\bibfnamefont {Z.}~\bibnamefont {Jiang}},\ and\ \bibinfo {author}
  {\bibfnamefont {R.}~\bibnamefont {Huan}},\ }\bibfield  {title} {\bibinfo
  {title} {Synchronization of electrically coupled micromechanical oscillators
  with a frequency ratio of 3:1},\ }\href {https://doi.org/10.1063/1.5000786}
  {\bibfield  {journal} {\bibinfo  {journal} {Applied Physics Letters}\
  }\textbf {\bibinfo {volume} {112}},\ \bibinfo {pages} {013503} (\bibinfo
  {year} {2018})},\ \Eprint
  {https://arxiv.org/abs/https://doi.org/10.1063/1.5000786}
  {https://doi.org/10.1063/1.5000786} \BibitemShut {NoStop}%
\bibitem [{\citenamefont {Matheny}\ \emph {et~al.}(2014)\citenamefont
  {Matheny}, \citenamefont {Grau}, \citenamefont {Villanueva}, \citenamefont
  {Karabalin}, \citenamefont {Cross},\ and\ \citenamefont
  {Roukes}}]{MathenyPRL}%
  \BibitemOpen
  \bibfield  {author} {\bibinfo {author} {\bibfnamefont {M.~H.}\ \bibnamefont
  {Matheny}}, \bibinfo {author} {\bibfnamefont {M.}~\bibnamefont {Grau}},
  \bibinfo {author} {\bibfnamefont {L.~G.}\ \bibnamefont {Villanueva}},
  \bibinfo {author} {\bibfnamefont {R.~B.}\ \bibnamefont {Karabalin}}, \bibinfo
  {author} {\bibfnamefont {M.~C.}\ \bibnamefont {Cross}},\ and\ \bibinfo
  {author} {\bibfnamefont {M.~L.}\ \bibnamefont {Roukes}},\ }\bibfield  {title}
  {\bibinfo {title} {Phase synchronization of two anharmonic nanomechanical
  oscillators},\ }\href {https://doi.org/10.1103/PhysRevLett.112.014101}
  {\bibfield  {journal} {\bibinfo  {journal} {Phys. Rev. Lett.}\ }\textbf
  {\bibinfo {volume} {112}},\ \bibinfo {pages} {014101} (\bibinfo {year}
  {2014})}\BibitemShut {NoStop}%
\bibitem [{\citenamefont {Karabalin}\ \emph {et~al.}(2009)\citenamefont
  {Karabalin}, \citenamefont {Cross},\ and\ \citenamefont
  {Roukes}}]{KarabalinPRB}%
  \BibitemOpen
  \bibfield  {author} {\bibinfo {author} {\bibfnamefont {R.~B.}\ \bibnamefont
  {Karabalin}}, \bibinfo {author} {\bibfnamefont {M.~C.}\ \bibnamefont
  {Cross}},\ and\ \bibinfo {author} {\bibfnamefont {M.~L.}\ \bibnamefont
  {Roukes}},\ }\bibfield  {title} {\bibinfo {title} {Nonlinear dynamics and
  chaos in two coupled nanomechanical resonators},\ }\href
  {https://doi.org/10.1103/PhysRevB.79.165309} {\bibfield  {journal} {\bibinfo
  {journal} {Phys. Rev. B}\ }\textbf {\bibinfo {volume} {79}},\ \bibinfo
  {pages} {165309} (\bibinfo {year} {2009})}\BibitemShut {NoStop}%
\bibitem [{\citenamefont {Ma}\ \emph {et~al.}(2014)\citenamefont {Ma},
  \citenamefont {You}, \citenamefont {Si}, \citenamefont {Xiong}, \citenamefont
  {Li}, \citenamefont {Yang},\ and\ \citenamefont {Wu}}]{PhysRevA.90.043839}%
  \BibitemOpen
  \bibfield  {author} {\bibinfo {author} {\bibfnamefont {J.}~\bibnamefont
  {Ma}}, \bibinfo {author} {\bibfnamefont {C.}~\bibnamefont {You}}, \bibinfo
  {author} {\bibfnamefont {L.-G.}\ \bibnamefont {Si}}, \bibinfo {author}
  {\bibfnamefont {H.}~\bibnamefont {Xiong}}, \bibinfo {author} {\bibfnamefont
  {J.}~\bibnamefont {Li}}, \bibinfo {author} {\bibfnamefont {X.}~\bibnamefont
  {Yang}},\ and\ \bibinfo {author} {\bibfnamefont {Y.}~\bibnamefont {Wu}},\
  }\bibfield  {title} {\bibinfo {title} {Formation and manipulation of
  optomechanical chaos via a bichromatic driving},\ }\href
  {https://doi.org/10.1103/PhysRevA.90.043839} {\bibfield  {journal} {\bibinfo
  {journal} {Phys. Rev. A}\ }\textbf {\bibinfo {volume} {90}},\ \bibinfo
  {pages} {043839} (\bibinfo {year} {2014})}\BibitemShut {NoStop}%
\bibitem [{\citenamefont {Larson}\ and\ \citenamefont
  {Horsdal}(2011)}]{PhysRevA.84.021804}%
  \BibitemOpen
  \bibfield  {author} {\bibinfo {author} {\bibfnamefont {J.}~\bibnamefont
  {Larson}}\ and\ \bibinfo {author} {\bibfnamefont {M.}~\bibnamefont
  {Horsdal}},\ }\bibfield  {title} {\bibinfo {title} {Photonic josephson
  effect, phase transitions, and chaos in optomechanical systems},\ }\href
  {https://doi.org/10.1103/PhysRevA.84.021804} {\bibfield  {journal} {\bibinfo
  {journal} {Phys. Rev. A}\ }\textbf {\bibinfo {volume} {84}},\ \bibinfo
  {pages} {021804} (\bibinfo {year} {2011})}\BibitemShut {NoStop}%
\bibitem [{\citenamefont {Jin}\ \emph {et~al.}(2017)\citenamefont {Jin},
  \citenamefont {Guo}, \citenamefont {Ji},\ and\ \citenamefont {Li}}]{Jin2017}%
  \BibitemOpen
  \bibfield  {author} {\bibinfo {author} {\bibfnamefont {L.}~\bibnamefont
  {Jin}}, \bibinfo {author} {\bibfnamefont {Y.}~\bibnamefont {Guo}}, \bibinfo
  {author} {\bibfnamefont {X.}~\bibnamefont {Ji}},\ and\ \bibinfo {author}
  {\bibfnamefont {L.}~\bibnamefont {Li}},\ }\bibfield  {title} {\bibinfo
  {title} {Reconfigurable chaos in electro-optomechanical system with negative
  duffing resonators},\ }\href {https://doi.org/10.1038/s41598-017-05020-w}
  {\bibfield  {journal} {\bibinfo  {journal} {Scientific Reports}\ }\textbf
  {\bibinfo {volume} {7}},\ \bibinfo {pages} {4822} (\bibinfo {year}
  {2017})}\BibitemShut {NoStop}%
\bibitem [{\citenamefont {Navarro-Urrios}\ \emph {et~al.}(2017)\citenamefont
  {Navarro-Urrios}, \citenamefont {Capuj}, \citenamefont {Colombano},
  \citenamefont {Garc{\'i}a}, \citenamefont {Sledzinska}, \citenamefont
  {Alzina}, \citenamefont {Griol}, \citenamefont {Mart{\'i}nez},\ and\
  \citenamefont {Sotomayor-Torres}}]{Navarro-Urrios2017}%
  \BibitemOpen
  \bibfield  {author} {\bibinfo {author} {\bibfnamefont {D.}~\bibnamefont
  {Navarro-Urrios}}, \bibinfo {author} {\bibfnamefont {N.~E.}\ \bibnamefont
  {Capuj}}, \bibinfo {author} {\bibfnamefont {M.~F.}\ \bibnamefont
  {Colombano}}, \bibinfo {author} {\bibfnamefont {P.~D.}\ \bibnamefont
  {Garc{\'i}a}}, \bibinfo {author} {\bibfnamefont {M.}~\bibnamefont
  {Sledzinska}}, \bibinfo {author} {\bibfnamefont {F.}~\bibnamefont {Alzina}},
  \bibinfo {author} {\bibfnamefont {A.}~\bibnamefont {Griol}}, \bibinfo
  {author} {\bibfnamefont {A.}~\bibnamefont {Mart{\'i}nez}},\ and\ \bibinfo
  {author} {\bibfnamefont {C.~M.}\ \bibnamefont {Sotomayor-Torres}},\
  }\bibfield  {title} {\bibinfo {title} {Nonlinear dynamics and chaos in an
  optomechanical beam},\ }\href {https://doi.org/10.1038/ncomms14965}
  {\bibfield  {journal} {\bibinfo  {journal} {Nature Communications}\ }\textbf
  {\bibinfo {volume} {8}},\ \bibinfo {pages} {14965} (\bibinfo {year}
  {2017})}\BibitemShut {NoStop}%
\bibitem [{\citenamefont {Wu}\ \emph {et~al.}(2017)\citenamefont {Wu},
  \citenamefont {Huang}, \citenamefont {Huang}, \citenamefont {Zhou},
  \citenamefont {Yang}, \citenamefont {Liu}, \citenamefont {Yu}, \citenamefont
  {Lo}, \citenamefont {Kwong}, \citenamefont {Duan},\ and\ \citenamefont
  {Wei~Wong}}]{Wu2017}%
  \BibitemOpen
  \bibfield  {author} {\bibinfo {author} {\bibfnamefont {J.}~\bibnamefont
  {Wu}}, \bibinfo {author} {\bibfnamefont {S.-W.}\ \bibnamefont {Huang}},
  \bibinfo {author} {\bibfnamefont {Y.}~\bibnamefont {Huang}}, \bibinfo
  {author} {\bibfnamefont {H.}~\bibnamefont {Zhou}}, \bibinfo {author}
  {\bibfnamefont {J.}~\bibnamefont {Yang}}, \bibinfo {author} {\bibfnamefont
  {J.-M.}\ \bibnamefont {Liu}}, \bibinfo {author} {\bibfnamefont
  {M.}~\bibnamefont {Yu}}, \bibinfo {author} {\bibfnamefont {G.}~\bibnamefont
  {Lo}}, \bibinfo {author} {\bibfnamefont {D.-L.}\ \bibnamefont {Kwong}},
  \bibinfo {author} {\bibfnamefont {S.}~\bibnamefont {Duan}},\ and\ \bibinfo
  {author} {\bibfnamefont {C.}~\bibnamefont {Wei~Wong}},\ }\bibfield  {title}
  {\bibinfo {title} {Mesoscopic chaos mediated by drude electron-hole plasma in
  silicon optomechanical oscillators},\ }\href
  {https://doi.org/10.1038/ncomms15570} {\bibfield  {journal} {\bibinfo
  {journal} {Nature Communications}\ }\textbf {\bibinfo {volume} {8}},\
  \bibinfo {pages} {15570} (\bibinfo {year} {2017})}\BibitemShut {NoStop}%
\bibitem [{\citenamefont {Roy}\ and\ \citenamefont
  {Devoret}(2016)}]{ROY2016740}%
  \BibitemOpen
  \bibfield  {author} {\bibinfo {author} {\bibfnamefont {A.}~\bibnamefont
  {Roy}}\ and\ \bibinfo {author} {\bibfnamefont {M.}~\bibnamefont {Devoret}},\
  }\bibfield  {title} {\bibinfo {title} {Introduction to parametric
  amplification of quantum signals with josephson circuits},\ }\href
  {https://doi.org/https://doi.org/10.1016/j.crhy.2016.07.012} {\bibfield
  {journal} {\bibinfo  {journal} {Comptes Rendus Physique}\ }\textbf {\bibinfo
  {volume} {17}},\ \bibinfo {pages} {740 } (\bibinfo {year} {2016})},\ \bibinfo
  {note} {quantum microwaves / Micro-ondes quantiques}\BibitemShut {NoStop}%
\bibitem [{\citenamefont {Hsuan}\ \emph {et~al.}(1967)\citenamefont {Hsuan},
  \citenamefont {Ajmera},\ and\ \citenamefont {Lonngren}}]{HsuanAPL}%
  \BibitemOpen
  \bibfield  {author} {\bibinfo {author} {\bibfnamefont {H.~C.~S.}\
  \bibnamefont {Hsuan}}, \bibinfo {author} {\bibfnamefont {R.~C.}\ \bibnamefont
  {Ajmera}},\ and\ \bibinfo {author} {\bibfnamefont {K.~E.}\ \bibnamefont
  {Lonngren}},\ }\bibfield  {title} {\bibinfo {title} {The nonlinear effect of
  altering the zeroth order density distribution of a plasma},\ }\href
  {https://doi.org/10.1063/1.1755133} {\bibfield  {journal} {\bibinfo
  {journal} {Applied Physics Letters}\ }\textbf {\bibinfo {volume} {11}},\
  \bibinfo {pages} {277} (\bibinfo {year} {1967})},\ \Eprint
  {https://arxiv.org/abs/https://doi.org/10.1063/1.1755133}
  {https://doi.org/10.1063/1.1755133} \BibitemShut {NoStop}%
\bibitem [{\citenamefont {Antoni}\ \emph {et~al.}(2011)\citenamefont {Antoni},
  \citenamefont {Kuhn}, \citenamefont {Briant}, \citenamefont {Cohadon},
  \citenamefont {Heidmann}, \citenamefont {Braive}, \citenamefont {Beveratos},
  \citenamefont {Abram}, \citenamefont {Gratiet}, \citenamefont {Sagnes},\ and\
  \citenamefont {Robert-Philip}}]{Antoni:11}%
  \BibitemOpen
  \bibfield  {author} {\bibinfo {author} {\bibfnamefont {T.}~\bibnamefont
  {Antoni}}, \bibinfo {author} {\bibfnamefont {A.~G.}\ \bibnamefont {Kuhn}},
  \bibinfo {author} {\bibfnamefont {T.}~\bibnamefont {Briant}}, \bibinfo
  {author} {\bibfnamefont {P.-F.}\ \bibnamefont {Cohadon}}, \bibinfo {author}
  {\bibfnamefont {A.}~\bibnamefont {Heidmann}}, \bibinfo {author}
  {\bibfnamefont {R.}~\bibnamefont {Braive}}, \bibinfo {author} {\bibfnamefont
  {A.}~\bibnamefont {Beveratos}}, \bibinfo {author} {\bibfnamefont
  {I.}~\bibnamefont {Abram}}, \bibinfo {author} {\bibfnamefont {L.~L.}\
  \bibnamefont {Gratiet}}, \bibinfo {author} {\bibfnamefont {I.}~\bibnamefont
  {Sagnes}},\ and\ \bibinfo {author} {\bibfnamefont {I.}~\bibnamefont
  {Robert-Philip}},\ }\bibfield  {title} {\bibinfo {title} {Deformable
  two-dimensional photonic crystal slab for cavity optomechanics},\ }\href
  {https://doi.org/10.1364/OL.36.003434} {\bibfield  {journal} {\bibinfo
  {journal} {Opt. Lett.}\ }\textbf {\bibinfo {volume} {36}},\ \bibinfo {pages}
  {3434} (\bibinfo {year} {2011})}\BibitemShut {NoStop}%
\bibitem [{\citenamefont {Chowdhury}\ \emph {et~al.}(2016)\citenamefont
  {Chowdhury}, \citenamefont {Yeo}, \citenamefont {Tsvirkun}, \citenamefont
  {Raineri}, \citenamefont {Beaudoin}, \citenamefont {Sagnes}, \citenamefont
  {Raj}, \citenamefont {Robert-Philip},\ and\ \citenamefont
  {Braive}}]{chowdhuryAPL}%
  \BibitemOpen
  \bibfield  {author} {\bibinfo {author} {\bibfnamefont {A.}~\bibnamefont
  {Chowdhury}}, \bibinfo {author} {\bibfnamefont {I.}~\bibnamefont {Yeo}},
  \bibinfo {author} {\bibfnamefont {V.}~\bibnamefont {Tsvirkun}}, \bibinfo
  {author} {\bibfnamefont {F.}~\bibnamefont {Raineri}}, \bibinfo {author}
  {\bibfnamefont {G.}~\bibnamefont {Beaudoin}}, \bibinfo {author}
  {\bibfnamefont {I.}~\bibnamefont {Sagnes}}, \bibinfo {author} {\bibfnamefont
  {R.}~\bibnamefont {Raj}}, \bibinfo {author} {\bibfnamefont {I.}~\bibnamefont
  {Robert-Philip}},\ and\ \bibinfo {author} {\bibfnamefont {R.}~\bibnamefont
  {Braive}},\ }\bibfield  {title} {\bibinfo {title} {Superharmonic resonances
  in a two-dimensional non-linear photonic-crystal nano-electro-mechanical
  oscillator},\ }\href {https://doi.org/10.1063/1.4947064} {\bibfield
  {journal} {\bibinfo  {journal} {Applied Physics Letters}\ }\textbf {\bibinfo
  {volume} {108}},\ \bibinfo {pages} {163102} (\bibinfo {year} {2016})},\
  \Eprint {https://arxiv.org/abs/https://doi.org/10.1063/1.4947064}
  {https://doi.org/10.1063/1.4947064} \BibitemShut {NoStop}%
\bibitem [{\citenamefont {Zanette}(2018)}]{Zanette_2018}%
  \BibitemOpen
  \bibfield  {author} {\bibinfo {author} {\bibfnamefont {D.~H.}\ \bibnamefont
  {Zanette}},\ }\bibfield  {title} {\bibinfo {title} {Energy exchange between
  coupled mechanical oscillators: linear regimes},\ }\href
  {https://doi.org/10.1088/2399-6528/aadfc6} {\bibfield  {journal} {\bibinfo
  {journal} {Journal of Physics Communications}\ }\textbf {\bibinfo {volume}
  {2}},\ \bibinfo {pages} {095015} (\bibinfo {year} {2018})}\BibitemShut
  {NoStop}%
\bibitem [{\citenamefont {Joe}\ \emph {et~al.}(2006)\citenamefont {Joe},
  \citenamefont {Satanin},\ and\ \citenamefont {Kim}}]{Joe_2006}%
  \BibitemOpen
  \bibfield  {author} {\bibinfo {author} {\bibfnamefont {Y.~S.}\ \bibnamefont
  {Joe}}, \bibinfo {author} {\bibfnamefont {A.~M.}\ \bibnamefont {Satanin}},\
  and\ \bibinfo {author} {\bibfnamefont {C.~S.}\ \bibnamefont {Kim}},\
  }\bibfield  {title} {\bibinfo {title} {Classical analogy of fano
  resonances},\ }\href {https://doi.org/10.1088/0031-8949/74/2/020} {\bibfield
  {journal} {\bibinfo  {journal} {Physica Scripta}\ }\textbf {\bibinfo {volume}
  {74}},\ \bibinfo {pages} {259} (\bibinfo {year} {2006})}\BibitemShut
  {NoStop}%
\bibitem [{\citenamefont {Limonov}\ \emph {et~al.}(2017)\citenamefont
  {Limonov}, \citenamefont {Rybin}, \citenamefont {Poddubny},\ and\
  \citenamefont {Kivshar}}]{Limonov2017}%
  \BibitemOpen
  \bibfield  {author} {\bibinfo {author} {\bibfnamefont {M.~F.}\ \bibnamefont
  {Limonov}}, \bibinfo {author} {\bibfnamefont {M.~V.}\ \bibnamefont {Rybin}},
  \bibinfo {author} {\bibfnamefont {A.~N.}\ \bibnamefont {Poddubny}},\ and\
  \bibinfo {author} {\bibfnamefont {Y.~S.}\ \bibnamefont {Kivshar}},\
  }\bibfield  {title} {\bibinfo {title} {Fano resonances in photonics},\ }\href
  {https://doi.org/10.1038/nphoton.2017.142} {\bibfield  {journal} {\bibinfo
  {journal} {Nature Photonics}\ }\textbf {\bibinfo {volume} {11}},\ \bibinfo
  {pages} {543} (\bibinfo {year} {2017})}\BibitemShut {NoStop}%
\bibitem [{\citenamefont {Stassi}\ \emph {et~al.}(2017)\citenamefont {Stassi},
  \citenamefont {Chiad{\`o}}, \citenamefont {Calafiore}, \citenamefont
  {Palmara}, \citenamefont {Cabrini},\ and\ \citenamefont
  {Ricciardi}}]{Stassi2017}%
  \BibitemOpen
  \bibfield  {author} {\bibinfo {author} {\bibfnamefont {S.}~\bibnamefont
  {Stassi}}, \bibinfo {author} {\bibfnamefont {A.}~\bibnamefont {Chiad{\`o}}},
  \bibinfo {author} {\bibfnamefont {G.}~\bibnamefont {Calafiore}}, \bibinfo
  {author} {\bibfnamefont {G.}~\bibnamefont {Palmara}}, \bibinfo {author}
  {\bibfnamefont {S.}~\bibnamefont {Cabrini}},\ and\ \bibinfo {author}
  {\bibfnamefont {C.}~\bibnamefont {Ricciardi}},\ }\bibfield  {title} {\bibinfo
  {title} {Experimental evidence of fano resonances in nanomechanical
  resonators},\ }\href {https://doi.org/10.1038/s41598-017-01147-y} {\bibfield
  {journal} {\bibinfo  {journal} {Scientific Reports}\ }\textbf {\bibinfo
  {volume} {7}},\ \bibinfo {pages} {1065} (\bibinfo {year} {2017})}\BibitemShut
  {NoStop}%
\bibitem [{\citenamefont {Zanotto}(2018)}]{Zanotto2018}%
  \BibitemOpen
  \bibfield  {author} {\bibinfo {author} {\bibfnamefont {S.}~\bibnamefont
  {Zanotto}},\ }\bibinfo {title} {Weak coupling, strong coupling, critical
  coupling and fano resonances: A unifying vision},\ in\ \href
  {https://doi.org/10.1007/978-3-319-99731-5_23} {\emph {\bibinfo {booktitle}
  {Fano Resonances in Optics and Microwaves: Physics and Applications}}},\
  \bibinfo {editor} {edited by\ \bibinfo {editor} {\bibfnamefont
  {E.}~\bibnamefont {Kamenetskii}}, \bibinfo {editor} {\bibfnamefont
  {A.}~\bibnamefont {Sadreev}},\ and\ \bibinfo {editor} {\bibfnamefont
  {A.}~\bibnamefont {Miroshnichenko}}}\ (\bibinfo  {publisher} {Springer
  International Publishing},\ \bibinfo {address} {Cham},\ \bibinfo {year}
  {2018})\ pp.\ \bibinfo {pages} {551--570}\BibitemShut {NoStop}%
\bibitem [{\citenamefont {Rieger}\ \emph {et~al.}(2012)\citenamefont {Rieger},
  \citenamefont {Faust}, \citenamefont {Seitner}, \citenamefont {Kotthaus},\
  and\ \citenamefont {Weig}}]{RiegerAPL}%
  \BibitemOpen
  \bibfield  {author} {\bibinfo {author} {\bibfnamefont {J.}~\bibnamefont
  {Rieger}}, \bibinfo {author} {\bibfnamefont {T.}~\bibnamefont {Faust}},
  \bibinfo {author} {\bibfnamefont {M.~J.}\ \bibnamefont {Seitner}}, \bibinfo
  {author} {\bibfnamefont {J.~P.}\ \bibnamefont {Kotthaus}},\ and\ \bibinfo
  {author} {\bibfnamefont {E.~M.}\ \bibnamefont {Weig}},\ }\bibfield  {title}
  {\bibinfo {title} {Frequency and q factor control of nanomechanical
  resonators},\ }\href {https://doi.org/10.1063/1.4751351} {\bibfield
  {journal} {\bibinfo  {journal} {Applied Physics Letters}\ }\textbf {\bibinfo
  {volume} {101}},\ \bibinfo {pages} {103110} (\bibinfo {year} {2012})},\
  \Eprint {https://arxiv.org/abs/https://doi.org/10.1063/1.4751351}
  {https://doi.org/10.1063/1.4751351} \BibitemShut {NoStop}%
\bibitem [{\citenamefont {Rhoads}\ \emph {et~al.}(2010)\citenamefont {Rhoads},
  \citenamefont {Shaw},\ and\ \citenamefont {Turner}}]{RhoadsJDS}%
  \BibitemOpen
  \bibfield  {author} {\bibinfo {author} {\bibfnamefont {J.~F.}\ \bibnamefont
  {Rhoads}}, \bibinfo {author} {\bibfnamefont {S.~W.}\ \bibnamefont {Shaw}},\
  and\ \bibinfo {author} {\bibfnamefont {K.~L.}\ \bibnamefont {Turner}},\
  }\bibfield  {title} {\bibinfo {title} {Nonlinear dynamics and its
  applications in micro- and nanoresonators},\ }\bibfield  {journal} {\bibinfo
  {journal} {Journal of Dynamic Systems, Measurement, and Control}\ }\textbf
  {\bibinfo {volume} {132}},\ \href {https://doi.org/10.1115/1.4001333}
  {10.1115/1.4001333} (\bibinfo {year} {2010}),\ \bibinfo {note} {034001},\
  \Eprint
  {https://arxiv.org/abs/https://asmedigitalcollection.asme.org/dynamicsystems/article-pdf/132/3/034001/5653925/034001\_1.pdf}
  {https://asmedigitalcollection.asme.org/dynamicsystems/article-pdf/132/3/034001/5653925/034001\_1.pdf}
  \BibitemShut {NoStop}%
\bibitem [{\citenamefont {Chowdhury}\ \emph {et~al.}(2019)\citenamefont
  {Chowdhury}, \citenamefont {Clerc}, \citenamefont {Barbay}, \citenamefont
  {Robert-Philip},\ and\ \citenamefont {Braive}}]{chowdhury2019weak}%
  \BibitemOpen
  \bibfield  {author} {\bibinfo {author} {\bibfnamefont {A.}~\bibnamefont
  {Chowdhury}}, \bibinfo {author} {\bibfnamefont {M.~G.}\ \bibnamefont
  {Clerc}}, \bibinfo {author} {\bibfnamefont {S.}~\bibnamefont {Barbay}},
  \bibinfo {author} {\bibfnamefont {I.}~\bibnamefont {Robert-Philip}},\ and\
  \bibinfo {author} {\bibfnamefont {R.}~\bibnamefont {Braive}},\ }\href@noop {}
  {\bibinfo {title} {Weak signal enhancement by non-linear resonance control in
  a forced nano-electromechanical resonator}} (\bibinfo {year} {2019}),\
  \Eprint {https://arxiv.org/abs/1910.04686} {arXiv:1910.04686
  [physics.app-ph]} \BibitemShut {NoStop}%
\bibitem [{\citenamefont {Hegger}\ \emph {et~al.}(1999)\citenamefont {Hegger},
  \citenamefont {Kantz},\ and\ \citenamefont {Schreiber}}]{Tisean}%
  \BibitemOpen
  \bibfield  {author} {\bibinfo {author} {\bibfnamefont {R.}~\bibnamefont
  {Hegger}}, \bibinfo {author} {\bibfnamefont {H.}~\bibnamefont {Kantz}},\ and\
  \bibinfo {author} {\bibfnamefont {T.}~\bibnamefont {Schreiber}},\ }\bibfield
  {title} {\bibinfo {title} {Practical implementation of nonlinear time series
  methods: The tisean package},\ }\href {https://doi.org/10.1063/1.166424}
  {\bibfield  {journal} {\bibinfo  {journal} {Chaos: An Interdisciplinary
  Journal of Nonlinear Science}\ }\textbf {\bibinfo {volume} {9}},\ \bibinfo
  {pages} {413} (\bibinfo {year} {1999})},\ \Eprint
  {https://arxiv.org/abs/https://doi.org/10.1063/1.166424}
  {https://doi.org/10.1063/1.166424} \BibitemShut {NoStop}%
\bibitem [{\citenamefont {Rosenstein}\ \emph {et~al.}(1993)\citenamefont
  {Rosenstein}, \citenamefont {Collins},\ and\ \citenamefont
  {Luca}}]{ROSENSTEIN1993117}%
  \BibitemOpen
  \bibfield  {author} {\bibinfo {author} {\bibfnamefont {M.~T.}\ \bibnamefont
  {Rosenstein}}, \bibinfo {author} {\bibfnamefont {J.~J.}\ \bibnamefont
  {Collins}},\ and\ \bibinfo {author} {\bibfnamefont {C.~J.~D.}\ \bibnamefont
  {Luca}},\ }\bibfield  {title} {\bibinfo {title} {A practical method for
  calculating largest lyapunov exponents from small data sets},\ }\href
  {https://doi.org/https://doi.org/10.1016/0167-2789(93)90009-P} {\bibfield
  {journal} {\bibinfo  {journal} {Physica D: Nonlinear Phenomena}\ }\textbf
  {\bibinfo {volume} {65}},\ \bibinfo {pages} {117 } (\bibinfo {year}
  {1993})}\BibitemShut {NoStop}%
\bibitem [{\citenamefont {Lee}\ \emph {et~al.}(1985)\citenamefont {Lee},
  \citenamefont {Yoon},\ and\ \citenamefont {Shin}}]{Shin1998}%
  \BibitemOpen
  \bibfield  {author} {\bibinfo {author} {\bibfnamefont {C.}~\bibnamefont
  {Lee}}, \bibinfo {author} {\bibfnamefont {T.}~\bibnamefont {Yoon}},\ and\
  \bibinfo {author} {\bibfnamefont {S.}~\bibnamefont {Shin}},\ }\bibfield
  {title} {\bibinfo {title} {Period doubling and chaos in a directly modulated
  laser diode},\ }\bibfield  {journal} {\bibinfo  {journal} {Applied Physics
  Letters}\ }\textbf {\bibinfo {volume} {46}},\ \href
  {https://doi.org/10.1063/1.95810} {10.1063/1.95810} (\bibinfo {year}
  {1985})\BibitemShut {NoStop}%
\bibitem [{\citenamefont {Cadeddu}\ \emph {et~al.}(2016)\citenamefont
  {Cadeddu}, \citenamefont {Braakman}, \citenamefont {T{\"u}t{\"u}nc{\"u}oglu},
  \citenamefont {Matteini}, \citenamefont {R{\"u}ffer}, \citenamefont
  {Fontcuberta~i Morral},\ and\ \citenamefont {Poggio}}]{Cadeddu2016}%
  \BibitemOpen
  \bibfield  {author} {\bibinfo {author} {\bibfnamefont {D.}~\bibnamefont
  {Cadeddu}}, \bibinfo {author} {\bibfnamefont {F.~R.}\ \bibnamefont
  {Braakman}}, \bibinfo {author} {\bibfnamefont {G.}~\bibnamefont
  {T{\"u}t{\"u}nc{\"u}oglu}}, \bibinfo {author} {\bibfnamefont
  {F.}~\bibnamefont {Matteini}}, \bibinfo {author} {\bibfnamefont
  {D.}~\bibnamefont {R{\"u}ffer}}, \bibinfo {author} {\bibfnamefont
  {A.}~\bibnamefont {Fontcuberta~i Morral}},\ and\ \bibinfo {author}
  {\bibfnamefont {M.}~\bibnamefont {Poggio}},\ }\bibfield  {title} {\bibinfo
  {title} {Time-resolved nonlinear coupling between orthogonal flexural modes
  of a pristine gaas nanowire},\ }\href
  {https://doi.org/10.1021/acs.nanolett.5b03822} {\bibfield  {journal}
  {\bibinfo  {journal} {Nano Letters}\ }\textbf {\bibinfo {volume} {16}},\
  \bibinfo {pages} {926} (\bibinfo {year} {2016})},\ \bibinfo {note} {pMID:
  26785132},\ \Eprint
  {https://arxiv.org/abs/https://doi.org/10.1021/acs.nanolett.5b03822}
  {https://doi.org/10.1021/acs.nanolett.5b03822} \BibitemShut {NoStop}%
\bibitem [{\citenamefont {Mercier~de L{\'e}pinay}\ \emph
  {et~al.}(2018)\citenamefont {Mercier~de L{\'e}pinay}, \citenamefont {Pigeau},
  \citenamefont {Besga},\ and\ \citenamefont {Arcizet}}]{MercierdeLepinay2018}%
  \BibitemOpen
  \bibfield  {author} {\bibinfo {author} {\bibfnamefont {L.}~\bibnamefont
  {Mercier~de L{\'e}pinay}}, \bibinfo {author} {\bibfnamefont {B.}~\bibnamefont
  {Pigeau}}, \bibinfo {author} {\bibfnamefont {B.}~\bibnamefont {Besga}},\ and\
  \bibinfo {author} {\bibfnamefont {O.}~\bibnamefont {Arcizet}},\ }\bibfield
  {title} {\bibinfo {title} {Eigenmode orthogonality breaking and anomalous
  dynamics in multimode nano-optomechanical systems under non-reciprocal
  coupling},\ }\href {https://doi.org/10.1038/s41467-018-03741-8} {\bibfield
  {journal} {\bibinfo  {journal} {Nature Communications}\ }\textbf {\bibinfo
  {volume} {9}},\ \bibinfo {pages} {1401} (\bibinfo {year} {2018})}\BibitemShut
  {NoStop}%
\bibitem [{\citenamefont {Boccaletti}\ \emph {et~al.}(2002)\citenamefont
  {Boccaletti}, \citenamefont {Kurths}, \citenamefont {Osipov}, \citenamefont
  {Valladares},\ and\ \citenamefont {Zhou}}]{BOCCALETTI20021}%
  \BibitemOpen
  \bibfield  {author} {\bibinfo {author} {\bibfnamefont {S.}~\bibnamefont
  {Boccaletti}}, \bibinfo {author} {\bibfnamefont {J.}~\bibnamefont {Kurths}},
  \bibinfo {author} {\bibfnamefont {G.}~\bibnamefont {Osipov}}, \bibinfo
  {author} {\bibfnamefont {D.}~\bibnamefont {Valladares}},\ and\ \bibinfo
  {author} {\bibfnamefont {C.}~\bibnamefont {Zhou}},\ }\bibfield  {title}
  {\bibinfo {title} {The synchronization of chaotic systems},\ }\href
  {https://doi.org/https://doi.org/10.1016/S0370-1573(02)00137-0} {\bibfield
  {journal} {\bibinfo  {journal} {Physics Reports}\ }\textbf {\bibinfo {volume}
  {366}},\ \bibinfo {pages} {1 } (\bibinfo {year} {2002})}\BibitemShut
  {NoStop}%
\end{thebibliography}
\end{document}